# Structural Brain Predictors of Visual Attention Gradient Modulated by Trait Anxiety

Suhail Rafiq Mir[1,4#], Dolcy Dhar[1], Ishita Singh[1], Prakrati Gupta[2], Varsha Singh[3#],

and Mrinmoy Chakrabarty[1#]

[1] Dept. of Social Sciences and Humanities, Indraprastha Institute of Information Technology Delhi, New Delhi 110020, INDIA.

[2] Dept. of Computer Science and Engineering, Indraprastha Institute of Information Technology Delhi, New Delhi 110020, INDIA.

[3] Dept. of Humanities and Social Sciences, Indian Institute of Technology Delhi, New Delhi 110016, INDIA.

[4] Dept. of Psychiatry, All India Institute of Medical Sciences, New Delhi 110029, INDIA.

[#] Equal contribution (Co-first author)

Correspondence (*):

Mrinmoy Chakrabarty, Ph.D.

Assistant Professor (Cognitive Science), Dept. of Social Sciences and Humanities, Indraprastha Institute of Information Technology Delhi (IIITD), New Delhi 110020, INDIA. TEL: +91-011-26907-363; E-mail: mrinmoy@iiitd.ac.in

Words in Main Text (excluding abstract, references and figure legends) = 9,576

Words in Abstract = 251

Figures = 4; Tables = 2

**Acknowledgments**

The authors thank the Centre for Advanced Research in Imaging, Neuroscience and Genomics (CARING) of Mahajan Imaging, SDA, New Delhi, Delhi 110016, for providing the MRI imaging facility and Ms. Madhuri Barnwal, Mr. Baby, and Mr. Sibin for scheduling and technical assistance during scanning. The authors are grateful to Dr. Sumitash Jana (IIT Delhi) and Dr. Jainendra Shukla (IIIT Delhi) for comments on parts of the manuscript, and also thank Mr. Anmol Shrivastav (Bhilai Institute of Technology; Summer Intern 2025 at IIIT Delhi) for assistance in the SBM analysis.

**Abstract**

Dynamic allocation of attention across the visual field, quantified as a visual-spatial attention gradient, is essential for maintaining perceptual breadth in visual cognition. Disruptions to this attentional flexibility may contribute to alterations in spatial attentional bias and could be influenced by affective traits, notably trait anxiety. In this study, we examined whether individual variability in structural brain features predicts differences in spatial attention deployment with regards to trait anxiety. Sixty participants deemed statistically adequate after a priori sample size calculation completed a visual-spatial attention gradient task incorporating brief, partial facial emotion cues. While no significant behavioural modulation of attention gradients by discrete emotional cues was observed, structural neuroimaging analyses revealed that increased grey matter volume in bilateral cerebellar lobule VI and greater cortical thickness in the left precentral gyrus and paracentral lobule were independently associated with reduced interaction between average attention gradient (across emotions; regardless of valence) and trait anxiety. Furthermore, machine learning models reliably predicted individual attention–anxiety profiles using these neuroanatomical features. These findings suggest that individuals with lower trait anxiety and greater structural integrity in cerebellar and sensorimotor regions exhibit more flexible attentional deployment. These results highlight the less known role of cerebellar and sensorimotor regions, traditionally associated with motor control, in modulating visual spatial attention, with regards to individual differences in trait anxiety. This work underscores a less reported neural mechanism in cognitive-affective regulation, revealing brain regions that support flexible attentional control in the face of anxiety, offering predictive value of structural brain markers for cognitive-affective interactions.

## 1.Introduction

Visuospatial attention enables efficient processing of visual stimuli, supporting perception and action in dynamic environments. Effective daily navigation requires flexible attention allocation across multiple spatial locations, such as searching for objects, crossing streets, or driving (Carrasco, 2011). Failure to distribute attention adaptively can impair visual field sampling and lead to attentional narrowing, hindering visually guided behaviour. Here, we conceptualize attentional narrowing as a continuous sharpening of the visuospatial attention gradient, reflecting disproportionate allocation of attentional resources to near locations at the expense of peripheral space (Robertson et al., 2013). Accordingly, steeper gradients are interpreted as greater severity of attentional narrowing.

This is particularly relevant in affective traits and disorders, such as anxiety, which distort visual processing. Anxiety worsens contrast sensitivity to fearful expressions (Ferneyhough *et al.*, 2013), heightens ratings of facial emotions (Li *et al.*, 2008), amplifies perceived contrast sensitivity (Barbot & Carrasco, 2018), and alters visual temporal resolution (Chakrabarty *et al.*, 2021). These visual processing effects extend to attentional biases. Anxiety leads to a tendency to focus on task-relevant or negative stimuli (Osinsky *et al.*, 2012), especially emotional facial expressions (Bradley *et al.*, 2000). Anxious individuals also show delayed disengagement from threatening cues (Fox *et al.*, 2001; Georgiou *et al.*, 2005). Similar biases are seen in patients with generalized anxiety disorder (GAD), who focus on threatening facial emotions (Bradley *et al.*, 1999; Mogg *et al.*, 2000). Evidence shows that individuals with high anxiety preferentially attend to threat-related stimuli (Mathews & MacLeod, 1994), with faster responses to threatening cues (Bradley *et al.*, 1999)and slower reactions to threat-related words in the emotional Stroop task (Mogg *et al.*, 1989). Cued attention studies reveal weaker spatial attention in anxious individuals at threat-related locations (Fox, 2002). Recent studies link trait anxiety to spatial attentional biases, particularly emotional cues (Dodd *et al.*, 2017; Kaur *et al.*, 2023).

While brain correlates of emotion-driven attentional biases have been explored using functional magnetic resonance imaging (MRI) and electroencephalography (EEG) (Vuilleumier & Driver, 2007; Fani *et al.*, 2012; Dominguez-Borras & Vuilleumier, 2013), the role of brain structure in the visual attention gradient, particularly concerning trait anxiety, remains unclear. Unlike functional neuroimaging, which captures state-dependent neural activations during task performance, the present study focuses on GMV and cortical thickness as trait-like neuroanatomical markers that index stable inter-individual differences. Structural measures are well suited to explain variability in attentional capacity and flexibility, as they reflect enduring properties of local neural architecture rather than transient task-evoked responses. We therefore adopted a morphometric approach to examine whether stable anatomical differences predict variability in the visual–spatial attention gradient and its modulation by trait anxiety. This study addresses this gap by examining how brain structural features predict the visual spatial attention gradient quantified as the difference in attentional efficiency between near and far locations relative to an emotional cue (Kaur *et al.*, 2023). A positive gradient indicates greater attention to the near location, while a negative gradient favours the far location. Here, spatial attention is operationalised as a graded distribution of attentional efficiency across the visual field rather than as a binary cue-validity effect. Specifically, we quantify how attentional efficiency varies as a function of spatial distance from a cue, yielding a continuous spatial attention gradient. The term *attention gradient* denotes a quantitative measure of how attentional efficiency varies as a function of spatial distance from a reference point, rather than a qualitative description of attentional focus. It reflects the relative prioritization of near versus far locations within the visual field. This gradient-based approach differs from classical Posner or dot-probe paradigms, which primarily index whether attention is allocated to a location, and instead captures how flexibly attention spreads across space (Robertson *et al.*, 2013; Kaur *et al.*, 2023).Critically, we examine whether trait anxiety modulates the relationship between regional brain structure and the steepness of visuospatial attention gradients. This interaction is grounded in converging evidence that stable affective dispositions influence how enduring structural brain characteristics are translated into functional biases in spatial attention, particularly along central-peripheral dimensions of visual processing. In a behavioural experiment combined with voxel- and surface-based morphometric analyses using structural MRI, we investigated: a) the effect of partial emotion cues on the attention gradient; b) the relationship between grey matter volume (GMV), cortical thickness, and the overall attention gradient (averaging across emotions); and c) the role of trait anxiety in these relationships. Our results showed GMV in Cerebellum lobule VI and cortical thickness in the Precentral Gyrus and Paracentral Lobule as independent predictors of the attention gradient, in conjunction with trait anxiety. However, briefly presented partial emotion cues did not significantly modulate the attention gradient.

## 2.Material and Methods

### 2.1 Participants

We recruited 60 young adults (23 females) after written informed consent. All were typically developing, with no reported neurological or psychiatric diagnoses and not on medication. They were naive to the study's aims and participated for remuneration. All were university students (post-secondary education) proficient in English and could complete self-reported questionnaires. The study was approved by the Institutional Review Board of Indraprastha Institute of Information Technology, Delhi, India vide letter reference ID: IRB-IIITD/FR/11/2022/03. Participant demographics are in Table 1. On Day 1, participants completed a questionnaire, a ~5-minute visual contrast thresholding task, and a ~70-minute behavioural experiment (Fig.1). On Day 2, they underwent a ~30-minute MRI session. The average interval between Day 1 and Day 2 across participants was 13.28 days (± 12.29).

### 2.2 Sample Size calculation

We determined the requisite sample size based on effect sizes reported in prior studies involving human participants that aligned with the objectives of our behavioural and MRI experiments. A behavioural study reported a significant association (Spearman's $\rho$ = –0.55) between trait anxiety and the modulation of visuospatial attention by negative affective pre-cues (Kaur *et al.*, 2022). These findings were consistent with an earlier study (Veerapa *et al.*, 2020), which observed a positive relationship between trait anxiety and overt attentional bias toward negative stimuli (Pearson's $r$ = 0.51; $R^2$ = 0.26). Given the limited literature directly linking brain structure to visual attention, we considered findings from a study (Bergmann *et al.*, 2016) close to our design, who reported significant correlations between visual working memory (VWM) performance and structural measures in primary visual cortex (V1), including GMV and cortical thickness (GMV Pearson's r = 0.512; cortical thickness Pearson's $r$ = 0.424). To ensure a conservative estimate for sample size, we used the smallest reported effect size ($r$ = 0.424) from all the above studies considered in an a priori power analysis using the ‘Exact’ method implemented in G*Power 3.1.9.7 (Faul *et al.*, 2009). Using a two-tailed alpha level of 0.05 and statistical power of 0.80, the calculation yielded a minimum required sample size of 41 participants to detect a significant correlation. Our final sample included 60 participants, exceeding this requirement and falling within the range of comparable peer-reviewed studies (Fox, 1996; Bishop, 2009; Chakrabarty *et al.*, 2021; Dhar *et al.*, 2024).

## 2.3 Subjective ratings (Questionnaire)

Participants completed the State–Trait Anxiety Inventory – STAI (Spielberger, 1983) to assess self-reported anxiety levels before the start of the behavioural experiment on Day 1. Trait anxiety indicates a relatively stable tendency to perceive a wide range of situations as threatening, and it serves as an index of an individual's general propensity to respond with anxiety to perceived stressors over time (Wiedemann, 2015; Spielberger, 2022). The trait subscale of the STAI consists of 20 items rated on a four-point Likert scale ranging from "Not at all" to "Very much so," yielding a total score between 20 and 80. Higher scores denote greater levels of trait anxiety. The measure demonstrates good temporal stability, with test–retest reliability coefficients reported between 0.69 and 0.89 over two months (Spielberger, C.D., Gorsuch, R.L., Lushene, R., Vagg, P.R., Jacobs, 1983). Although both state and trait anxiety were assessed to characterize the sample, our analyses focused on trait anxiety because the study aimed to associate stable individual differences in visuospatial attention gradients to structural brain measures. State anxiety, which primarily reflects transient emotional arousal, was not entered into inferential models as it is less suited to explaining trait-like structure–behaviour associations.

## 2.4 Contrast Threshold of Visual Target

Each participant's target contrast threshold (just-noticeable-difference) was adaptively estimated at the farthest / largest eccentricity ($6^{o}$) using responses collected over a block of 50 trials implemented in PsychoPy v2022.1.3. This procedure was designed to determine the contrast level corresponding to 80% accuracy at the largest target eccentricity from the centre of the screen, which was then used throughout the experiment. Threshold estimation followed a modified one-up, three-down staircase procedure: contrast increased after each incorrect response and decreased following three consecutive correct responses. The step size was initially set at 8 dB and was reduced after every two staircase reversals, reaching a minimum of 1 dB. The individual threshold was defined as the mean contrast level across the final six reversals. This procedure ensured participants were equated at baseline regarding task difficulty, thereby controlling for inter-individual differences in visual sensitivity. Thresholding at the most demanding spatial location (farthest / largest eccentricity (6°) ensured perceptual visibility without ceiling effects while preserving sensitivity to attentional modulation. This allowed for the isolation of the effects of emotional facial cues on visuospatial attention, consistent with the study's hypothesis. The average contrast threshold across participants was 0.17 (SD ± 0.13).

**2.5 Behavioural Experiment**

Participants were seated 57 cm from a 27-inch ACER monitor (1920 × 1080 pixels; 144 Hz; subtending ~60 × 34° of visual angle), with head and chin rests used to stabilize gaze. Visual stimuli were presented to participants in a well-lighted room using custom scripts written in PsychoPy (v2021.2.1). Each trial commenced with a central white fixation cross (1 × 1°), displayed randomly between 300 and 500 ms. The offset of the fixation cross was immediately followed by the simultaneous presentation of a spatial attention cue (a filled black circle, 0.5° in diameter, positioned 5° left or right of centre on the horizontal meridian) and a central image prime (7 × 7/2° visual angle) for 50 ms. The spatial cue was intended to direct covert attention toward one visual hemifield in preparation for the upcoming target. At the same time, the image prime served as a task-irrelevant emotional or neutral stimulus. Following a 150 ms inter-stimulus interval (ISI), the target stimulus—a Gabor patch (1° diameter; spatial frequency = 3 cycles/degree; contrast = at individual participant's visual contrast threshold; luminance = 28.9 cd/m²) was presented for ~40 ms at one of three vertical eccentricities (±1.5°, 3°, or 6° from the cue location), randomly on either the left or right side of the display. A distractor stimulus (a black open circle; 1° diameter; contrast = 100%) was simultaneously presented at the location diametrically opposite the target. After stimulus presentation, an instruction screen prompted participants to respond using a standard PC keypad within 2 s. Each trial ended upon response within 2s, after which visual feedback was provided, followed by a 500 ms blank interval before the subsequent trial. The inter-stimulus interval (ISI) was fixed across all trials and conditions and implemented as 21 screen refresh frames on a 144 Hz display (21 × 6.9 ms = 144.9 ms), reported as ~150 ms for convenience in Figure 1A. To prevent saccades and isolate covert attentional processes, the interval between the onset of the image prime + spatial cue and the target was fixed at 200 ms, well under the latency of typical eye movements (~300 ms). Participants were instructed to look at a central fixation throughout each trial, attend covertly to the peripheral cue, ignore the central image prime, and report the orientation (left- or right-tilt) of the Gabor target as quickly and accurately as possible. Participants reported the orientation of a briefly presented Gabor patch rather than performing a simple detection or speeded response task. We chose this since simple reaction time measures may conflate attentional orienting with motor preparation and decisional urgency, all of which are systematically influenced by anxiety and arousal (Bishop, 2009). Orientation discrimination on the other hand, requires visual encoding under attentional constraints and therefore, provides a more direct behavioural index of attentional modulation. Accordingly, orientation sensitivity was used here as a validated readout of spatial attentional allocation. Emotional evaluation of the primes was not required, as the task focused on perceptual discrimination under conditions of explicit awareness of the primes but without directing attention to them.

Each participant completed six experimental sessions (72 trials per session; total = 432 trials) in a single lab visit. Within each session, 24 trials were primed with Neutral, Fearful faces, and Scrambled (control) images. For both emotional face conditions, primes consisted of an equal number of male and female faces (12 each). Each image type was presented equally across the three vertical eccentricities (1.5°, 3°, and 6°). Trial order was randomized within each session to eliminate the predictability of prime type and target location. The critical manipulation was the unexpected presentation of task-irrelevant image primes (Neutral, Fear, or Scrambled human faces) preceding the task-relevant target. If emotional content modulated spatial attention, attentional efficiency—quantified as performance across target eccentricities—would differ between emotion-bearing and control conditions. Before the main experiment, participants completed 10–20 practice trials, and the task began only after verbal confirmation that they fully understood the instructions. The entire session lasted approximately 70 minutes per participant.

Seventy-two unique grayscale face images (36 male, 36 female) were used, comprising equal numbers of Neutral and Fearful expressions. Images were sourced from validated, non-commercial databases: the Karolinska Directed Emotional Faces (Lundqvist, D., Flykt, A., Ohman, 1998) KDEF emotion ratings reported by Goeleven et al. (2008) were: valence = 6.23 (Fear), 4.75 (Neutral); arousal = 3.71 (Fear), 2.67 (Neutral). All images were converted to grayscale and luminance-normalized (mean pixel intensity = 128; range = 0–255) following procedures described by (Chakrabarty & Wada, 2020). Scrambled images, used as control primes, were generated by random pixel permutation of grayscale face images to eliminate facial structure while preserving overall luminance and contrast. Next, a facial landmark detection model was utilised to computationally identify salient facial features corresponding to the eyebrows and eyes, implemented via the OpenCV library (Gary Bradski, 2000). Pupil coordinates were estimated by computing the intersection of four key landmark points surrounding each, following the facial landmark schema explained elsewhere (Kazemi & Sullivan, 2014). Finally, face images were cropped using OpenCV so that only the eye and eyebrow regions were visible, and all images were resized. Each unique face image per emotion was shown twice across different sessions per participant. Fearful eye-region images were specifically chosen because the eyes and eyebrows alone convey reliable threat-related emotional information and have been shown to modulate attention and perception even in the absence of full facial context (Whalen *et al.*, 2001; Sadr *et al.*, 2003). Their use allowed us to test whether a minimal, task-irrelevant threat signal could provide the affective perturbation to bias visuospatial attention gradients. The distance between the eye and eyebrow was not manipulated but reflected natural variation across facial identities and expressions. The white scleral region of the eye was not independently quantified but only the eye-eyebrow region was treated as a subtle emotional cue. Figure 1 illustrates the experimental paradigm.

**2.6 Structural Magnetic Resonance Imaging (MRI) Experiment**

The anatomical images of the participants' brains were collected at the Magnetic Resonance Imaging facility of Mahajan Imaging & Labs, New Delhi, INDIA. The T1-weighted anatomical images were obtained from all the participants with the following acquisition parameters: Field Strength = 3 Tesla, 3D Ax T1 MP Rage, Repetition time/TR = 2.4s, Echo time/TE =2.3ms, Inversion time = 1s, Voxel size = 1.0 ×1.0 ×1.0, Matrix size = 256 × 256, Flip angle = 25°, Slice thickness = 1mm, Number of slices = 176. The present study employed voxel-based morphometry (VBM) and surface-based morphometry (SBM) to quantify regional gray matter volume and cortical thickness, respectively. These measures captured local gray-matter architecture rather than white-matter connectivity. While diffusion-based techniques are optimal for assessing tract integrity, our questions concerned regional structural substrates associated with individual differences in attentional deployment. Accordingly, GMV and cortical thickness were treated as complementary indices of local computational capacity within nodes of established cognitive networks.

### 2.7. 1 Behavioural Analysis

Keypad response data were analysed using custom-written scripts. The analysis procedure for a single participant is outlined below. Within each session, responses were sorted by emotion condition (Neutral, Fear, Scrambled) and target eccentricity (1.5°, 3°, and 6°) within each emotion. For each condition, the median reaction time (RT) from correct trials, those in which the participant accurately discriminated the target orientation and the proportion of correct responses (accuracy) were used to compute the **negative inverse efficiency score (nIES)**, an index of attentional efficiency (Kaur *et al.*, 2023). This metric is a modified version of the inverse efficiency score originally proposed earlier (Townsend & Ashby, 1978), wherein higher nIES values indicate greater attentional efficiency. The inverse efficiency score combines RT and accuracy into a single metric, allowing a robust assessment of speed-accuracy trade-offs commonly used in attentional studies (Townsend & Ashby, 1978; Robertson *et al.*, 2013). The nIES was calculated separately for each of the three eccentricities and across all three emotion conditions. To account for session-level variability, the nIES scores for each of the six sessions were then independently standardised (z-scored) across sessions for each emotion condition. Note that the nIES values were calculated at three spatial locations relative to the central emotional cue presented at the screen centre: Near (1.5°), Intermediate (3°), and Far (6°). These were included to characterise the spatial distribution of emotion-cued attentional modulation. However, the Intermediate eccentricity (3°) was excluded from subsequent statistical analyses. This exclusion was intentional and theory-driven. The visuospatial attention gradient was defined as the difference between Near and Far locations to maximize spatial separation and yield a single, interpretable index of global attentional bias across space. The intermediate eccentricity served to ensure dense spatial sampling and to confirm monotonic changes in attentional efficiency, but was not included in the gradient computation to avoid capturing local effects, consistent with prior gradient-based approaches (Carrasco, 2011; Robertson *et al.*, 2013; Kaur *et al.*, 2023).

To quantify the spatial gradient of visuospatial attention (hereafter referred to as the 'gradient') in each participant, a central-to-peripheral attentional gradient was computed in each emotion as the difference in normalized inverse efficiency scores (nIES) between 1.5° and 6° eccentricity. This directly quantified the magnitude of attentional fall-off across the maximal spatial separation in the task design, with 1.5° indexing near-central processing and 6° indexing far-peripheral processing. The resulting difference score therefore, captured the steepness and direction of spatial attentional bias and the extent of peripheral attenuation. Given that emotion conditions did not significantly modulate this directional gradient, subsequent MRI analyses focused on individual differences in the overall strength of spatial attentional modulation independent of sign as follows.

A participant-level composite index of spatial attentional narrowing was derived by combining the emotion-specific gradients using the Euclidean norm across conditions. The Euclidean norm (i.e., the square root of the summed squared gradients) provides a magnitude-based aggregation of multiple components in multidimensional space (L2 norm), preserving overall gradient strength while preventing directional cancellation and ensuring equal weighting across the experimental conditions, e.g., emotional conditions (Strang, 2009). Conceptually, this procedure yields the root-mean-square magnitude of spatial attentional gradients across emotional conditions, thereby indexing the average strength of spatial attentional bias independent of sign. The resulting scalar composite gradient was entered as a regressor in the voxel- and surface-based morphometry analyses. This metric captures the steepness or overall strength of spatial attentional modulation irrespective of whether attention was biased toward near or far space, and is therefore, more appropriate for relating stable structural brain features to individual variability in attentional flexibility. This endpoint-based metric captures global spatial prioritization across the visual field rather than local variations and is mathematically equivalent to a slope estimate between the two extremes. Larger values indicate steeper gradients, corresponding to stronger spatial attentional bias across eccentricities or spatially restricted attention, whereas attenuated gradients indicate more uniform attentional deployment. The Composite Gradient therefore, provides a quantitative index of the magnitude (strength) of spatial attentional bias irrespective of direction, such that larger values reflect greater differentiation in attentional allocation across spatial locations, whereas attenuated gradients indicate broader and more flexible attentional deployment. The spatial attention gradient was computed as a within-subject difference between Near and Far target locations, reducing the influence of global factors such as anxiety-related tonic arousal. This derived gradient score was used in subsequent statistical analyses. The average task accuracy across all six sessions was > 80 **%** per participant. The group-level mean accuracy was **89 ± 0.09% (mean ± SD)**.

### 2.7.2 Voxel-Based Morphometry (VBM) Analysis

All imaging data were processed and analyzed using Statistical Parametric Mapping -SPM12, version 7771, (Penny *et al.*, 2007) in conjunction with the Computational Anatomy Toolbox (CAT12, version 12.8.1, r2043) (Gaser *et al.*, 2020), implemented in MATLAB 2023a (MathWorks Inc.). Structural T1-weighted scans were initially aligned to a common anatomical space such that the anterior commissure of each image coincided with that of the default SPM template. Image preprocessing followed the standard CAT12 segmentation pipeline, employing the Adaptive Maximum a Posteriori (AMAP) method, which does not rely on pre-defined tissue priors. Initial segmentation divided the brain into gray matter (GM), white matter (WM), and cerebrospinal fluid (CSF), followed by Partial Volume Estimation (PVE) to account for mixed tissue classes (GM-WM and GM-CSF). Before segmentation, all images underwent affine registration to a standard template, skull stripping, and normalization to Montreal Neurological Institute (MNI) space. Bias-corrected, segmented GM images were subjected to CAT12's sample homogeneity quality check. Individual Image Quality Ratings (IQRs) were recorded, and only scans exceeding a threshold of 70% were retained for analysis (mean ± SD = 86.07 ± 4.47, range: 73.12 – 93.02). Total intracranial volume (TIV) was estimated for each participant (mean ± SD = 1.45 ± 0.15 $cm^3$). Finally, GM tissue probability maps were smoothed with an 8 mm full-width at half-maximum (FWHM) Gaussian kernel. This voxel-based morphometry (VBM) approach enables voxel-wise estimation of regional gray matter volume (rGMV) at a spatial resolution of 1× 1× $1mm^3$, allowing for precise statistical analysis of structural brain differences across cortical and subcortical regions. Smoothed gray matter images were entered into a second-level multiple regression analysis, with the composite attention-gradient magnitude (Euclidean norm across conditions) and its interaction with trait anxiety specified as the covariates of interest, while controlling for age and TIV. This analysis aimed to identify associations of gradient and gradient in the context of trait anxiety with regional gray matter morphology. Trait anxiety was not included as a nuisance covariate in the voxel-wise morphometric analyses because statistically treating trait anxiety as a nuisance regressor would remove variance shared with attentional performance and exclude the focal relationship of interest, i.e., anxiety-dependent modulation of attentional gradients, while imposing an additive model inappropriate for conceptually interdependent predictors (Miller & Chapman, 2001; Yzerbyt *et al.*, 2004). The independent variables in the model were orthogonal. To identify significant differences in regional GMV, a voxel-level height threshold of $p < 0.001$ (uncorrected) was applied across the whole brain. To control for multiple comparisons and reduce the risk of false positives, we subsequently applied family-wise error (FWE) correction at the voxel level, with a significance threshold of $p < 0.05$ (FWE-corrected). When voxel-level FWE correction did not yield significant results, correction was applied at the cluster level, using a significance threshold of $p < 0.05$ (FWE-corrected). Only clusters or voxels surviving this correction were considered statistically significant. Anatomical labelling of significant regions was done using the CAT12_AAL3 atlas (Tzourio-Mazoyer *et al.*, 2002). Additionally, voxel-wise GMVs were extracted from the above significant regions and subjected to linear regression and predictive modelling analyses.

### 2.7.3 Surface-Based Morphometry (SBM) Analysis

T1-weighted structural images were analysed using the CAT12 and SPM12 toolboxes implemented in MATLAB2023a to perform surface-based morphometry (SBM). Cortical surface reconstruction followed the standard SBM pipeline within the CAT12 framework. This included cortical thickness estimation and central surface reconstruction using the projection-based thickness (PBT) method. Topological correction was applied following surface reconstruction, and the resulting surfaces were further refined to produce the final central surface mesh. Individual central surfaces were registered to the Freesurfer 'FsAverage' template via spherical mapping. Subsequently, surface meshes for all participants were smoothed using a 15 mm full-width at half-maximum (FWHM) Gaussian kernel, in line with previous work (Lerch & Evans, 2005) and recommendation (Gaser *et al.*, 2020). The independent variables in the model were orthogonal. Subsequently, vertex-wise correlations between cortical thickness and the composite attention-gradient magnitude (Euclidean norm across conditions) and its interaction with trait anxiety were computed across the cortical surface. Trait anxiety was not included as a nuisance covariate in the vertex-wise morphometric analyses because of the same reason as explained under 'VBM analysis' above. The resulting statistical maps were thresholded at a vertex-level of $p < 0.005$ (uncorrected) across the whole cortical surface, a threshold previously demonstrated to offer a sensitive and appropriate balance between Type I and Type II error rates with 15mm FWHM in cortical thickness analysis (Lerch & Evans, 2005). To control for multiple comparisons and reduce the likelihood of false positives, a cluster-level family-wise error (FWE) correction was subsequently applied, with a significance threshold of $p < 0.05$ (FWE-corrected). Only clusters or vertices surviving this correction were deemed statistically significant. Anatomical labelling of significant regions was performed using the CAT12_AAL3 atlas (Tzourio-Mazoyer *et al.*, 2002). Consistent with the VBM analysis, vertex-wise cortical thickness values were extracted from the aforementioned significant regions and subsequently entered into linear regression and predictive modelling analyses.

**2.7.4 Prediction analysis**

We employed a machine learning approach to identify brain-based predictors of individual variability in attention gradient as followed earlier (Qin *et al.*, 2014). Data from significant clusters in whole-brain VBM and SBM analyses were further analysed using a machine learning approach (5-fold cross-validation + linear regression) to identify neuroanatomical predictors of individual differences in the gradient regarding anxiety (i.e., gradient × trait anxiety interaction). Attention gradient served as the dependent variable, and structural brain measures (total GMV of bilateral cerebellum lobule VI and total cortical thickness of left precentral gyrus and paracentral lobule in two independent analyses) were used as the independent variable. In the first step, the model's predictive accuracy was assessed by computing the correlation between predicted and observed values [Pearson's *r*(predicted, observed)]. The dataset was partitioned into five folds, ensuring that the distributions of the dependent and independent variables were balanced across folds. A linear regression model was trained on four folds and tested on the remaining fold. This process was iterated five times, with each fold used once as the test set. The resulting *r*(predicted, observed) represented the average model performance across folds. A nonparametric permutation test was conducted to evaluate statistical significance. An empirical null distribution of *r*(predicted, observed) was generated by permuting the target variable labels in 2,000 surrogate datasets under the null hypothesis of no association between attention gradient and cerebellar volume, on the one hand, and cortical thickness of left precentral gyrus and paracentral lobule, on the other hand. For each surrogate dataset, *r*(predicted, observed)_surrogate was computed using the same 5-fold cross-validation procedure. The *p*-value was derived by calculating the proportion of permuted *r*(predicted, observed)_surrogate values exceeding the observed / empirical *r*(predicted, observed). Note that only the hypothesized relationships between attention gradient, trait anxiety, and regional morphometry (GMV or cortical thickness) were examined. No additional models were tested or omitted.

**[Insert Figure 1]**

## 3.Results

### 3.1 Effects of partial Emotion pre-cues on Attention gradient

A preliminary analysis revealed a significant effect of eccentricity on attentional efficiency ($F(2,118) = 13.56$, $p = 0.000035$, partial $\eta^2 = 0.19$), with attentional efficiency decreasing as target eccentricity increased from near (1.5°) to far (6°) locations across emotion conditions (see Supplementary Information – Analysis 1; Fig. 2). Having established this eccentricity-dependent attentional gradient, the main investigation was to determine whether image primes conveying different emotional signals modulated visuospatial attention gradients. Consequently, a one-way repeated-measures ANOVA was conducted with emotion condition as a within-subject factor (Neutral, Fear, Scrambled). Assumptions of normality (Lilliefors test; all $p$s $\geq 0.16$) and sphericity (Mauchly’s test: $W = 0.99$, $p = 0.72$) were confirmed. This revealed no significant main effect of emotion condition on gradients ($F_{(2, 118)} = 0.19$, $p = 0.82$, partial $\eta^2 = 0.0033$; Supplementary Fig. 1), indicating that brief, partial emotion cues did not systematically bias the spatial allocation of visual attention. Given the absence of emotion-specific modulation of the signed gradient, subsequent brain-behaviour analyses focused on the single composite gradient magnitude (Euclidean norm) per participant, reflecting the overall strength of spatial attentional modulation independent of directional bias. Although the Euclidean norm is direction-invariant, examination of the signed Near minus Far gradients used to derive the final Euclidean norm-based metric revealed that approximately 75% of participants exhibited positive gradients (Supplementary Fig.2). Thus, near-space prioritization represented the predominant attentional profile in the present sample, even though the Euclidean norm itself indexes the magnitude of spatial attentional bias irrespective of direction.

To facilitate the interpretation of the subsequent neuroimaging analyses, we additionally examined the direct associations between anxiety measures and behavioural attention indices. Trait anxiety was not significantly associated with nIES values at any eccentricity or emotion condition, nor with the composite gradient measure. Likewise, trait and state anxiety were not significantly correlated, and state anxiety was not significantly associated with the composite gradient measure. Further, whole-brain analyses examining trait anxiety as the sole predictor of gray matter volume and cortical thickness revealed no significant FWE-corrected effects, indicating an absence of direct morphometric correlates of trait anxiety in the present sample (all $p > .05$; see Supplementary Information Analysis 2; Table 1 for complete results). Additional analyses examining the reliability of the gradient measure, its convergence with an alternative three-eccentricity slope-based gradient metric, and the robustness of the neuroimaging findings are reported in the supporting Supplementary Analysis 3.

**[Insert Figure 2]**

### 3.2 Relationship of GMV with Attention Gradient and Trait Anxiety

To investigate how the gradient, independently and in interaction with trait anxiety, relates to whole-brain GMV, we performed multiple regression analyses after controlling for age and total intracranial volume (TIV) as covariates. The interaction term was computed by multiplying each participant's gradient score by their corresponding trait anxiety score.

We first assessed the association between the whole-brain GMV and gradient. Voxel-wise multiple regression identified a significant negative correlation (Pearson's $r_{(58)} = -0.40$, 95% confidence interval = [−0.59, −0.16]) in the right cerebellar lobule VI (whole-brain voxel-wise FWE correction at $p$ <0.05;- $T_{(56)} = 5.16$, cluster-level $pFWE = 0.03$, cluster size $kE = 4$; Table 2-A, Fig. 3A-B). Specifically, larger GMV in this region corresponded to reduced gradient values, indicative of greater regional grey matter architecture associated with enhanced flexibility in visuospatial attentional deployment. No other region showed positive/negative correlations.

Second, we explored how the interaction between gradient and trait anxiety influences GMV, aiming to clarify whether the relationship between gradient and grey matter varies across individuals with differing levels of trait anxiety. Using voxel-wise regression incorporating the interaction term (attention gradient × trait anxiety), we identified a significant effect in bilateral cerebellar lobule VI (cluster-defining threshold of $p <$ 0.001 (uncorrected) and cluster-extent threshold = 80 expected voxels per cluster, considered significant at a cluster-level FWE-corrected threshold of $p < 0.05$; $T_{(56)} = 3.24$, cluster-level $pFWE < 0.022$, cluster size $kE \geq$ 871 voxels ; see Table 2-B and Fig. 3C–D). Notably, the previously observed negative correlation between GMV and gradient in this region was further modulated by trait anxiety severity indicating that individuals with higher anxiety showed a stronger inverse GMV- gradient relationship (slope = -1.63, Pearson's $r_{(28)}$ = -0.43, 95% confidence interval = [−0.68, −0.08]), whereas this effect was attenuated in those with lower anxiety (slope = -1.01, Pearson's $r_{(28)}$ = -0.28, 95% confidence interval = [−0.58, 0.09]). This suggests that trait anxiety moderates how bilateral cerebellum lobule VI relates to visuospatial attentional flexibility. To clarify the nature of the interaction, the observed effects do not reflect a simple additive strengthening of gray matter volume-attention associations driven by anxiety-related structural differences. Rather, trait anxiety selectively amplified the coupling between regional GMV and attentional gradients within cerebellar regions. Lower anxiety did not reverse this relationship, nor did anxiety independently account for additional structural variance. Instead, anxiety moderated the slope of the GMV-attention gradient association, such that structure-attention coupling emerged preferentially at higher levels of trait anxiety.

Next, we stratified the data by gender to uncover underlying patterns. The data was split separately by the median trait anxiety scores of females (median = 47) and males (median = 42) into High Anxiety subset (≥ median, $n_{females}$ = 12, $n_{males}$ = 21) and Low Anxiety subset (< median, $n_{females}$ = 11, $n_{males}$ = 16) respectively. We observed a notable divergence in the relationship between GMV, gradient, and trait anxiety: this modulation effect of trait anxiety on the GMV–gradient association appeared stronger in females (High Anxiety subset: slope = -2.85, Pearson's $r_{(10)}$ = -0.79, 95% confidence interval = [− 0.89, − 0.60]; Low Anxiety subset: slope = -2.78, Pearson's $r_{(9)}$ = -0.66, 95% confidence interval = [− 0.82 , − 0.39]) than in males (High Anxiety subset: slope = -1.00, Pearson's $r_{(19)}$ = -0.23, 95% confidence interval = [− 0.55, 0.14]; Low Anxiety subset: slope = -1.73, Pearson's $r_{(14)}$ = -0.28, 95% confidence interval = [− 0.58, - 0.09]; see Supplementary Fig. 3). However, when we statistically compared the slopes of these gender-specific trends using analysis of covariance (ANCOVA) in each anxiety subset, the difference did not reach significance (High Anxiety subset: $F_{(1,29)}$ = 2.27, $p$ = 0.14; Low Anxiety subset: $F_{(1,23)}$ = 0.28, $p$ = 0.59), likely reflecting the limited sample size within each gender subset. Further, there were no differences of either gradient (independent $t_{(58)}$ = 0.53, $p$ = 0.59) or trait anxiety (independent $t_{(58)}$ = -1.23, $p$ = 0.22) between males and females.

We then implemented a machine-learning approach using linear regression with balanced five-fold cross-validation to test whether total GMV in bilateral cerebellar lobule VI could predict individual gradient × trait-anxiety interaction scores. The model demonstrated significant predictive validity, $r$(predicted, observed) = -0.49, two-tailed $p$ = 0.009 (Fig. 3E–F), indicating that greater bilateral lobule VI volume reliably predicts lower gradient values, reflecting enhanced attentional flexibility, particularly when accounting for trait anxiety severity.

Overall, individuals with a lesser GMV of cerebellar lobule VI manifested greater magnitudes of spatial attentional bias, which was more pronounced in females than in males. Cerebellar lobule VI may interact with trait anxiety in tuning fine-grained spatial attention, which may predict individual differences in the magnitude of spatial attentional bias.

**[Insert Figure 3]**

### 3.3 Relationship of Cortical Thickness with Attention Gradient and Trait Anxiety

To examine how the gradient—independently and in interaction with trait anxiety—is related to whole-brain cortical thickness, we conducted multiple regression analyses, controlling for age. The interaction term was calculated by multiplying each participant's gradient score by their corresponding trait anxiety score as earlier. This approach allowed us to assess the main effects and the moderating role of trait anxiety on cortical thickness.

First, we examined the relationship between whole-brain cortical thickness and the gradient, which revealed significant negative correlation in the cluster consisting of the regions: left inferior frontal gyrus and left precentral gyrus (Pearson's $r_{(58)} = -0.33$, 95% confidence interval =[-0.54, -0.08]) which was statistically significant(cluster-defining threshold of $p < 0.005$ (uncorrected) and cluster-extent threshold = 111 vertices expected per cluster, considered significant at a cluster-level FWE-corrected threshold of $p < 0.05$; $T_{(57)} = 3.89$, cluster-level $pFWE = 0.024$, cluster size $kE = 526$ vertices; see Table 2-C; Fig. 4A–B). These results indicate that greater cortical thickness in lateral prefrontal and premotor regions is associated with lower gradient values, suggesting enhanced flexibility in visuospatial attentional deployment. No additional brain regions exhibited significant positive or negative correlations.

Subsequently, we examined whether the interaction between gradient and trait anxiety influenced cortical thickness—specifically, whether trait anxiety moderated the relationship between gradient and cortical thickness across individuals. Whole-brain regression revealed significant effects in the cluster consisting regions from left precentral gyrus and left paracentral lobule (cluster-defining threshold of $p <$ 0.005 (uncorrected) and cluster-extent threshold = 111 expected vertices per cluster, considered significant at a cluster-level FWE-corrected threshold of $p < 0.05$; $T_{(57)} = 3.41$, cluster-level $pFWE = 0.049$, cluster size $kE =$ 446 vertices; Table 2-D; Fig 4C–D) Mirroring our GMV results, the negative correlation between cortical thickness and gradient in these regions was significantly modulated by trait anxiety. Individuals higher in trait anxiety displayed a stronger inverse relationship (slope = -0.52, Pearson's $r_{(28)}$ = -0.51, 95% confidence interval = [− 0.73, − 0.17]), i.e., thicker cortex linked to enhanced attentional flexibility, whereas this effect was diminished among those lower in trait anxiety (slope = -0.18, Pearson's $r_{(28)}$ = -0.21, 95% confidence interval = [− 0.52, 0.17]). These findings suggest that trait anxiety significantly moderates the structural basis of visuospatial attentional flexibility in motor regions. The pattern of this interaction was similar to the VBM analysis. The interaction indicates that higher trait anxiety increased the sensitivity of cortical thickness-attention gradient coupling, particularly within sensorimotor regions. This effect reflects a conditional amplification of the cortical thickness-attention relationship under higher anxiety. Trait anxiety thus, functioned as a moderator that gated the expression of cortical thickness influences on attentional gradients.

When stratified by gender, we observed a faint relationship among cortical thickness, gradient, and trait anxiety in the data. Specifically, the modulatory impact of trait anxiety on the association between cortical thickness and the gradient was comparable between the two genders after running an ANCOVA ($F_{(1,56)} = 0.19$, $p = 0.67$; see Supplementary Fig. 4). Please note that we did not median split the data into High and Low Anxiety subsets as there was no apparent effect between the genders after inspecting the scatter plots.

Finally, we applied a machine-learning approach, using linear regression with five-fold cross-validation, to evaluate whether total cortical thickness in the left precentral gyrus and left paracentral lobule could predict each participant's gradient × trait anxiety interaction score. The model demonstrated significant predictive accuracy, $r$(predicted, observed) = -0.44, two-tailed $p = 0.026$ (Fig. 4E–F). This finding indicates that greater cortical thickness in these motor-related regions reliably predicts lower gradient values reflecting enhanced attentional flexibility—particularly when accounting for variations in trait anxiety severity.

**[Insert Figure 4]**

**4.Discussion**

In the present study, we examined whether brief emotional signals conveyed by partial facial expressions modulate the gradient of visual spatial attention. Unlike classical dot-probe or Posner paradigms that yield binary cue-validity effects, our task characterises spatial attention as a continuous gradient across eccentricity. This approach is particularly suited to examining attentional narrowing and individual differences related to trait anxiety, which may not be fully captured by simple reaction time benefits at cued locations. We further explored whole-brain grey matter volume (GMV) and cortical thickness correlates independently associated with the overall gradient (averaged across the trials primed by three partial emotional cues) in the context of inter-individual differences of trait anxiety. Although no significant attentional bias was observed in response to partial facial emotion cues, structural neuroimaging revealed that GMV in bilateral cerebellar lobule VI, as well as total cortical thickness in the left (inferior frontal gyrus, precentral gyrus, and paracentral lobule), were independently and negatively associated with the gradient. Additionally, GMV in bilateral cerebellar lobule VI and cortical thickness in the left precentral and paracentral lobules were negatively associated with the interaction term between the attention gradient and trait anxiety. These findings suggest that individual differences in structural brain substrates may contribute to the modulation of spatial attention, contingent upon trait anxiety severity. Specifically, greater GMV and cortical thickness were associated with attenuated gradients — indicative of more flexible spatial attention deployment in individuals with lower trait anxiety. Though gender differences in this relationship were apparent with GMV, they were not statistically significant, likely due to the limited gender-wise sample size. Complementary machine-learning analyses reliably predicted the gradient–trait anxiety interaction term using these structural metrics, supporting their dissociable roles. In this context, the observed associations between structural brain measures and the visuospatial attention gradient can be interpreted in terms of individual differences in spatial attentional bias. Individuals with lower cerebellar lobule VI volume or reduced sensorimotor cortical thickness exhibited steeper gradients, reflecting greater attentional narrowing (greater attentional differentiation) across eccentricities, whereas greater structural integrity predicted attenuated gradients and broader, flexible attentional deployment across space. Collectively, these results underscore the unique contributions of regional GMV and cortical thickness in shaping individual variability in visual attention deployment concerning the affective trait of anxiety.

There are reports of partial face information, e.g., the eye region and the eyebrows, sufficient for conveying emotions for recognition (Sadr *et al.*, 2003). Therefore, in one of the study's goals, we explored whether a subtle emotion signal conveyed by the average distance between the two eyes and eyebrow regions of human face images presented briefly as pre-cues was sufficient to elicit biases of attention distribution in space. We did not observe any biases in attentional deployment by the pre-cues. Although group-level performance showed the expected decline with eccentricity, individual gradients exhibited substantial variability, including negative values. In this framework, negative gradients indicate reduced prioritization of near space versus peripheral. Such idiosyncratic patterns have been reported previously and reflect individual differences in attentional breadth and spatial prioritization rather than measurement unreliability (Greenwood & Parasuraman, 1999; Robertson *et al.*, 2013; Kaur *et al.*, 2023). There could be a few reasons for the null effects in our results. We presented the images briefly (50 ms) to sufficiently convey the emotion signals, as opposed to the no time limit used earlier (Sadr *et al.*, 2003). We also excluded the forehead region of the faces from the pre-cues, which were found to be informative by Ekman and colleagues (Ekman & Friesen, 1978), and the exclusion of the forehead regions may deplete the information content of the pre-cues to bias attentional deployment in space. Moreover, emotional biases on one visual function may not generalize to others. For example, while fearful face cues can rapidly capture attention and enhance visual acuity under certain conditions, such emotional effects do not necessarily improve performance in more fine-grained visual tasks such as shape or colour discrimination (Bocanegra & Zeelenberg, 2011; Del Zotto & Pegna, 2015; Shang *et al.*, 2021). So, it is possible that at least partial emotion cues comprising the eyes and eyebrows, as implemented by us, do not impact gradients in healthy young adults. In our findings, one or more of the above reasons could have led to the null effects of emotional cues on gradient.

We observed that the GMV of bilateral cerebellum lobule VI predicts the interaction between gradient and trait anxiety. The negative association suggests that cerebellum lobule VI confers greater flexibility in attentional deployment, particularly in individuals with lower levels of trait anxiety. *Although cerebellar and sensorimotor regions are classically associated with motor control, converging evidence demonstrates their* *involvement in covert spatial attention and predictive attentional timing independent of overt movement. In the present task, short cue–target intervals were used to minimize eye movements, and the attention gradient was derived from perceptual efficiency rather than motor output, supporting an attentional interpretation of these structural associations.* Our findings linking the structural characteristics of cerebellar lobule VI to the visual spatial attention gradient align with recent fMRI evidence indicating functional organization in neighbouring cerebellar regions. Specifically, it was reported that cerebellar lobule VIIb/VIIIa, an area anatomically close to our region of concern being separated by the Crus I/II regions (Lobule VIIa), exhibits an organization similar to the parietal cortex (Brissenden *et al.*, 2018). Their study demonstrated that the dorsomedial portion of this lobule encodes spatial locations of attention, while the ventromedial portion is selectively engaged by increasing working memory demands. These results led the authors to conclude that the cerebellar representation of the visual field parallels the topographic mapping previously documented in the parietal and frontal regions of the cerebral cortex. This generally agrees with an earlier study, which also reported disturbances of overt shift of visual attention in cerebellar patients (Golla *et al.*, 2005). Traditionally, the cerebellum has been recognized for its role in motor control (Strick *et al.*, 2009). However, recent studies have illuminated its involvement in higher-order cognitive, social, and emotional processes. The cerebellum contributes to executive functions such as planning, attention, cognitive flexibility (Bellebaum & Daum, 2007), and interpreting goal-directed actions in social contexts (Van Overwalle *et al.*, 2019). Emerging evidence also suggests that structural anomalies of the cerebellum may be related to the interplay between the affective trait of depression and the severity of clinical symptoms in Autism Spectrum Disorder (Dhar *et al.*, 2024). Notably, the GMV of cerebellar lobule VI is positively correlated with the spontaneity and correct rate in a 3-back working memory task (Ding *et al.*, 2012). Elsewhere, it was reported that the activation of the cerebellum lobule VI scaled linearly by the load of the items in the working memory (Kirschen *et al.*, 2005). Furthermore, EEG evidence has established a context-specific causal role of the cerebellum in modulating anticipatory neural dynamics underlying temporal prediction, which in turn supports the proactive allocation of attentional resources and preparation of adaptive responses (Breska & Ivry, 2020). Additionally, precise symptom-lesion mapping of attention deficits has localized the lesions linked with abnormal reaction times in a covert visual attention task to cerebellum lobule VI, amongst others (Baier *et al.*, 2010). An earlier study has evidenced double dissociation wherein attention activated an area within lobule VI, and a motor task activated a distinct area within the anterior lobe – lobule III–IV (Allen *et al.*, 1997). Converging evidence from VBM, lesion, functional imaging, and brain stimulation studies demonstrates that cerebellar GMV, particularly within the posterior lobules is critically involved in the rapid and precise allocation of visuospatial attention. This involvement positions the cerebellum as an integral component of the broader cortical attentional network. Here, we extend the evidence for the cerebellum's extra-motor roles by demonstrating that volumetric variation in a cerebellar sub-region (lobule VI) predicts individual differences in visual spatial-attention deployment, which in turn correlates with trait-anxiety severity - an association not previously documented.

Structural variability in GMV within well-documented, visuospatial-relevant cortical regions consistently emerges as a fundamental correlate of higher visuospatial cognitive functions. Studies have demonstrated that greater GMV in occipitoparietal areas—including the lateral occipital complex (LOC) and intraparietal sulcus (IPS)—predicts superior visual working memory and spatial attention performance. For instance, individuals exhibiting larger left LOC and bilateral superior IPS volumes showed significantly higher capacity and precision in visual working memory paradigms, underscoring a direct relationship between cortical GMV and representational fidelity in visuospatial tasks (Luck & Vogel, 2013; Fukuda *et al.*, 2015). The IPS of the dorsal visual stream, in particular, has been repeatedly implicated in visuospatial attention and working memory, with volume differences in this region correlating with individual variability in visual short-term memory capacity, which emphasizes the region's map-based architecture and its role in visual spatial cognition (Dimond *et al.*, 2019). Complementary studies have also highlighted the role of angular gyrus in allocating visuospatial attention, where volumetric enhancements likely refine not only spatial orienting and dynamic attention shifts (Cattaneo *et al.*, 2009) but also being causally responsible for coordinate transformation while spatial updating (Muggleton *et al.*, 2008) and guiding visual spatial attention in reaching a decision based on the reward-punishment contingencies of the environment (Studer *et al.*, 2014). These findings agree with functional lesion and neglect studies pointing to IPS and posterior parietal cortex as critical hubs for spatial orienting and attentional selection (Corbetta & Shulman, 2011; Ptak & Schnider, 2011).

Significantly, emerging neuroimaging evidence now extends this dorsal attention framework to include the cerebellum, particularly its posterior lobules, which exhibit robust functional connectivity with core dorsal attention network (DAN) nodes. Specifically, lobules VI and VII (including Crus I and II) demonstrate synchronized activity with the IPS and frontal eye fields during goal-directed attentional tasks, implicating cerebellar sub-regions in modulating the precision and timing of attentional allocation (Buckner, 2013; Brissenden *et al.*, 2016). These cerebellar lobules show domain-specific recruitment during visuospatial processing and visual working memory, reinforcing that the cerebellum contributes through sensorimotor prediction and attentional control mechanisms. Moreover, intrinsic functional connectivity studies reveal a reproducible topography within the cerebellum that mirrors the cortical organization of large-scale networks, including the DAN, further supporting the cerebellum's role in attentional dynamics (Buckner *et al.,* 2011; Marek *et al.*, 2018). As such, individual variability in cerebellar GMV, particularly in DAN-linked lobules, may parallel the cortical structural correlates of visuospatial cognition, refining our understanding of distributed network interactions that underlie attention and working memory. While these insights have illuminated the cerebellum's role in domain-general attentional processes, our findings uniquely extend this framework by demonstrating that individual differences in **cerebellar GMV** specifically in lobule VI not only predict the gradient of visuospatial attention but are also **significantly modulated by trait anxiety**, a dimension of affect that has not been previously integrated into this structural-attentional mapping. This interaction between cerebellar morphology and anxiety underscores a **previously unreported affective modulation** of cerebellar contributions to attention, revealing a critical affect-cognition interface within the dorsal attention network. By uncovering this structural substrate of anxiety-linked variability in attention deployment, our study offers a novel account of how the cerebellum may support **adaptive or maladaptive tuning of attentional gradients** under emotional dispositions.

Our surface-based morphometry analysis also revealed the cortical thickness of the region comprising the left inferior frontal gyrus, precentral gyrus, and paracentral lobule to predict the interindividual differences of gradient in the context of trait anxiety. The current findings offer insights into the neuroanatomy underpinning visuospatial attentional gradients, particularly in the context of individual differences in trait anxiety. Specifically, we observed that cortical thickness in the left inferior temporal gyrus and precentral gyrus negatively correlated with attention gradient scores, indicating that thicker cortices in these regions are associated with more flexible spatial attention deployment. Notably, we found the involvement of the left inferior frontal gyrus (IFG), a region implicated in cognitive control, top-down modulation of attention, and resolution of attentional conflict (Aron *et al.*, 2004; Derrfuss *et al.*, 2004; Hampshire *et al.*, 2010). The IFG has been shown to interact with sensorimotor systems to guide goal-directed behaviour, especially while suppressing irrelevant stimuli or task switching - functions highly relevant to flexible visuospatial attention (Asplund *et al.*, 2010; Higo *et al.*, 2011). Its structural involvement in our findings suggests that greater cortical thickness in the IFG may enhance attentional adaptability, possibly by enabling more effective top-down control over motor-preparatory systems during attentional shifts. These results align with prior research linking motor and premotor regions to attentional control processes, particularly under task demands involving spatial orientation and voluntary attentional shifts (Corbetta *et al.*, 2008; Smith, 2015). Crucially, we extend this literature by demonstrating that trait anxiety significantly moderates this relationship, such that individuals higher in anxiety show a stronger inverse association between cortical thickness and gradient. This suggests that in high-anxiety individuals, motor-associated cortical regions may play a more prominent role in supporting attentional flexibility—possibly as a compensatory mechanism to counteract anxiety-related attentional narrowing (Bishop, 2009; Sylvester *et al.*, 2012). While previous studies have associated trait anxiety with functional alterations in attentional control networks - dorsolateral prefrontal cortex, anterior cingulate cortex (Bishop, 2007; Etkin *et al.*, 2009), our findings are among the first to implicate structural cortical differences—specifically in precentral and paracentral regions in anxiety-linked modulation of visuospatial attention gradients. Moreover, the observation that this modulation occurs in sensorimotor areas reinforces emerging models proposing that attentional control engages motor-preparatory systems even in covert tasks (Rushworth *et al.*, 2001; Moore & Fallah, 2004). The cortical regions identified by the gradient-only and gradient × trait anxiety analyses were anatomically adjacent yet non-overlapping, a pattern consistent with the distinct processes captured by each analysis: the former indexing affect-independent attentional architecture, and the latter isolating anxiety-specific modulation of attentional control. Taken together, these findings highlight a novel affect-structural interaction, wherein trait anxiety dynamically influences the relationship between cortical architecture and attentional flexibility, offering new avenues for understanding the neural basis of individual variability in cognitive-affective integration. It’s noteworthy however, that the cortical thickness interaction finding should be interpreted cautiously. While the effect survived the predefined statistical threshold, it was comparatively closer to the significance threshold and therefore, may be more sensitive to sampling variability. Replication in larger independent cohorts will be important to establish the robustness and generalizability of this association. Our present findings offer convergent evidence that GMV and cortical thickness represent structurally dissociable but functionally complementary substrates underpinning individual variability in visuospatial attention, especially in relation to trait anxiety. Gray matter volume, particularly in bilateral cerebellar lobule VI, was found to predict the interaction between trait anxiety and attentional gradient negatively — an association that aligns with emerging models of cerebellar involvement in higher-order cognitive and affective processes (Stoodley & Schmahmann, 2009; Buckner, 2013). Broadly, GMV has been interpreted as reflecting cellular density, dendritic arborization, and glial support (Panizzon *et al.*, 2009), and volumetric enhancement in cerebellar lobule VI has been associated with more efficient working memory, attentional timing, and proactive cognitive control (Kirschen *et al.*, 2005; Ding *et al.*, 2012; Breska & Ivry, 2020). Alongside, cortical thickness in the left precentral gyrus and paracentral lobule — regions implicated in voluntary orienting and motor-preparatory aspects of attention-was also negatively associated with attention gradients, particularly among individuals with higher anxiety levels. Cortical thickness, which typically reflects intracortical cytoarchitectural maturity including dendritic complexity and synaptic density (Narr *et al.*, 2007; Fjell & Walhovd, 2010), has been linked to top-down attentional modulation and flexibility during goal-directed tasks (Moore & Fallah, 2004; Corbetta *et al.*, 2008). Notably, the dissociation between GMV and cortical thickness in predicting attention gradient suggests that while cerebellar volume may govern predictive timing and spatial encoding of attention shifts, sensorimotor cortical thickness may support the voluntary reorientation and dynamic updating of attention in affectively challenging contexts. Together, the structural integrity of these neurobiological substrates may interact to optimize more adaptive, fine-grained deployment of visual-spatial attention, with their influence modulated by dispositional affective traits such as anxiety. Here, the term “structural integrity” refers specifically to local gray-matter characteristics indexed by GMV and cortical thickness, rather than to white-matter tract coherence or inter-regional connectivity. This structure-function-behaviour link offers explanatory insights into how distributed neuroanatomical features, particularly within cerebellar and premotor circuits, may alleviate or amplify trait anxiety-related constraints on controlling visual attention across points in space.

Finally, we identify a few aspects of the present study that may be considered for interpreting our present results and for the future. First, microsaccades, small eye movements, may have influenced covert attentional shifts (Gu *et al.*, 2024), though this remains speculative without eye-tracking data. Second, gender-related trends in the relationship between structural brain metrics and attentional gradient did not reach statistical significance, likely due to limited gender-specific sample sizes. Gender differences in anxiety-related neurocognitive manifestations (McLean *et al.*, 2011) warrant further exploration in larger, gender-balanced samples. Third, although our sample was adequately powered, a larger sample size would help validate these findings. Fourth, although clusters in the SBM analysis were identified using a cluster-defining threshold of $p < 0.005$ (uncorrected), supported by sensitivity analyses in a prior study (Lerch & Evans, 2005), confirming these findings with a more stringent threshold of $p < 0.001$ would be preferable. Fifth, the overlap between trait anxiety and depression (Knowles & Olatunji, 2020) may mean our results may reflect broader negative affectivity, which future studies should explore more explicitly. Sixth, while state anxiety primarily modulates transient perceptual gain and emotional response amplitude, particularly under conditions of threat, the present findings highlight trait anxiety as a stable affective disposition that interacts with cerebellar and sensorimotor structure to shape baseline visuospatial attention gradients. Future work may explicitly examine the contribution of state anxiety within this framework. Seventh, although the present study primarily used an endpoint-based Near–Far gradient metric, supplementary analyses showed that the principal findings were highly consistent with an alternative three-eccentricity linear slope-based measure, supporting the robustness of the reported results across different methods of quantifying attentional gradients. *Eighth,* future studies incorporating diffusion tensor imaging or tract-based morphometry would be valuable for directly assessing white-matter connectivity and cerebello-cortical communication, thereby complementing the present gray-matter-focused findings. Finally, because the present study employed a correlational design, the observed relationships between regional brain morphology, trait anxiety, and visuospatial attention gradients should be interpreted as associations rather than evidence of causal or mechanistic influence. Future studies will be necessary to determine the directionality and causal basis of these structure-behaviour relationships.

In conclusion, our study provides evidence that individual differences in the structural architecture of both cerebellar and cortical regions predict variability in visuospatial attentional deployment as a function of trait anxiety. While partial facial emotion cues failed to elicit spatial attentional biases, volumetric and cortical thickness analyses revealed dissociable yet complementary contributions of bilateral cerebellar lobule VI and left inferior frontal, precentral–paracentral regions in modulating attention gradients. Specifically, greater cerebellar GMV and sensorimotor cortical thickness are associated with attenuated gradients, particularly in individuals with lower anxiety levels, suggesting more flexible allocation of spatial attention. These findings extend existing models of the dorsal attention network by incorporating cerebellar regions into affectively modulated attentional control, highlighting a previously less-known structural mechanism through which anxiety influences visuospatial attention. By further integrating surface-based morphometry and machine learning with trait-level affective data, this study offers a framework for understanding how cerebellar and cortical morphometry may mutually shape cognitive-affective functions. Future studies incorporating eye-tracking, gender-balanced samples, and refined affective measures are essential to clarify the specificity of these findings and their translational relevance to anxiety-linked attentional dysfunctions. Overall, our results agree with the emerging yet underreported cerebellar-cortical structural substrate for individual variability in attention control under emotional dispositions, opening new avenues for personalized interventions targeting anxiety-related attentional dysregulation.

**Author contributions**

Conceptualization: Mrinmoy Chakrabarty, Varsha Singh; Methodology: Suhail Rafiq Mir, Dolcy Dhar, Prakrati Gutpa, Ishita Singh, Mrinmoy Chakrabarty; Formal analysis and investigation: Mrinmoy Chakrabarty, Suhail Rafiq Mir, Ishita Singh, Varsha Singh; Writing: Mrinmoy Chakrabarty, Suhail Rafiq Mir,

Varsha Singh; Funding acquisition: Mrinmoy Chakrabarty, Varsha Singh; Supervision: Mrinmoy Chakrabarty, Varsha Singh. All authors read and approved the final manuscript.

**Conflicts of Interest**

The authors declare no conflicts of interest.

**Funding**

The research was funded by the Multi-Institute Faculty Interdisciplinary Research Project grant (MFIRP-222) jointly awarded by the Indian Institute of Technology Delhi and the Indraprastha Institute of Information Technology Delhi to Dr. Mrinmoy Chakrabarty and Dr. Varsha Singh as well as the Science and Engineering Research Board Core Research Grant awarded to Dr. Mrinmoy Chakrabarty (SERB-CRG/2022/008119). Dr Suhail Rafiq Mir was supported by a Post-doctoral Fellowship from the Centre for Design and New Media (A TCS Foundation Initiative supported by the Tata Consultancy Services) at IIIT-Delhi. The funders were not involved in the study design, collection, analysis, interpretation of data, the writing of this article, or the decision to submit it for publication.

**Data availability:** The data and codes that led to the reported results can be accessed here : https://github.com/research-archive/Attention-Gradient

**Table 1: Demographic Information of the Participants**

| **Particulars** | **Ratio/Mean ± Std. deviation (Range)** |
|---|---|
| Gender ratio (males: females) | 37:23 (ratio) |
| Age in years | 22.8±3.8 (18-31) |
| STAI-Trait | 48.78±5.20 (37-65) |
| STAI-State | 43.52±5.19 (28-57) |

*Note.* STAI State Trait Anxiety Inventory

**Table 2**
**Results of brain-wide statistical analyses**

| Analysis | Brain Region | Peak MNI Coordinates | | | Cluster Size (# voxels, A &B; # vertices, C & D) | Cluster-level *p*-FWE |
|---|---|---|---|---|---|---|
| | | X | Y | Z | | |
| A) rGMV vs Attention gradient regression | Right Cerebellum Lobule VI | 33 | -56 | -28 | 7 | 0.027 |
| B) rGMV vs (Attention gradient x Trait anxiety interaction term regression | Right Cerebellum Lobule VI | 33 | -57 | -28 | 1964 | 0.001 |
| | Left Cerebellum Lobule VI | -21 | -60 | -26 | 871 | 0.022 |
| C) Cortical thickness vs Attention gradient regression | Frontal Inf Tri Left | -40 | 23 | 23 | 535 | 0.022 |
| | Left Precentral Gyrus | -45 | -1 | 36 | | |
| D) Cortical thickness vs (Attention gradient x Trait anxiety interaction term) regression | Left Precentral Gyrus | -23 | -24 | 66 | 446 | 0.049 |
| | Left Paracentral Lobule | -13 | -21 | 72 | | |
| | Left Precentral Gyrus | -40 | -20 | 59 | | |

*Note.* MNI Montreal Neurological Institute, *p*FWE Family-Wise Error corrected *p*-value. All *p*-values reported are family-wise error corrected (p-FWE), as detailed in the Methods.

**Figure 1. Behavioural task design**. (A) The timeline of the events of one trial of the behavioural task. All temporal parameters were fixed. The inter-stimulus interval (ISI) was programmed as 21 frames on a 144 Hz monitor (1 frame = 6.9 ms), corresponding to 144.9 ms and therefore shown as ~150 ms for schematic clarity. (B) The different image primes convey partial facial emotion Information (Fear, FE; Neutral, NE; Scrambled/Null). Arrows indicate the vertical distance from the centre of the pupil to the eyebrow; this feature was illustrative and not experimentally manipulated. (C) The three vertical eccentricities/spatial positions (±1.5, 3, 6 degrees) were measured from the imaginary position (black, open circle; 5 degrees from the screen centre) of the attention cue presentation on either half of the screen. The targets were randomly presented at each eccentricity, either above/below or left/right side, relative to the screen centre. The distractor was always presented at a position diametrically opposite to the target (180 degrees).

**Figure 2. Effects of emotions on the visuospatial attention gradient.** Plots showing negative inverse efficiency scores (nIES) at three spatial target eccentricities for an individual participant (line and marker) in the Neutral (NE, black circles), Fear (FE, blue triangles), and Scrambled (Scr, magenta squares) conditions.

**Figure 3. Relationship of GMV with Attention Gradient and Trait Anxiety.** (A) Negative correlation between attention gradient and GMV in the right cerebellar lobule VI. (B) A scatter plot shows the correlation in (A). (C) Correlation between the interaction term of attention gradient × trait anxiety and bilateral cerebellar GMV. (D) A scatter plot illustrates the correlation in (C). (E) Scatter plot illustrating the relationship between rGMV in bilateral cerebellum lobule VI and the width of spatial attention modulated by trait anxiety severity. The width measure was operationalized as the inverse of the Euclidean norm-based composite attention gradient (Width = $1/\|gradient\|_2$), such that larger values indicate broader spatial attention (shallower gradients). Each circular marker represents one participant. Marker size is scaled proportionally to the magnitude of the width measure ($1/\|gradient\|_2$), with larger markers indicating broader attentional spread / more flexible attention. Marker colour and the colour bar encode trait anxiety severity. The figure visualizes how cerebellar GMV predicts the width of attention, and how this relationship is modulated by trait anxiety severity. (F) Probability density plot of the null distribution of Pearson's correlation coefficients (*r*) from 2,000

random permutations, estimating the null hypothesis that the empirical (*r*) does not differ from zero for significance testing. The red dashed vertical line represents the empirical cross-validated (*r*). Thresholded statistical maps along with their respective colour bars are visualised using the MNI 152 standard template in MRIcroGL for (A) and (C). Least squares trend lines (broken black, red [high anxiety subset], and blue [low anxiety subset]) are shown in (B) and (D). Prediction intervals (dashed black lines) are shown in E.

**Figure 4. Relationship of cortical thickness with Attention Gradient and Trait Anxiety.**

Negative correlation between attention gradient and cortical thickness in the left (inferior frontal gyrus + precentral gyrus). (B) A scatter plot shows the correlation in (A). (C) Correlation between the interaction term of attention gradient × trait anxiety and the cortical thickness of the left (precentral gyrus + paracentral lobule). (D) A scatter plot illustrates the correlation in (C). (E) Scatter plot illustrating the relationship between cortical thickness of the left (precentral gyrus + paracentral lobule) and the width of spatial attention modulated by trait anxiety severity. The width measure was operationalized as the inverse of the Euclidean norm-based composite attention gradient (Width = $1/\|\text{gradient}\|_2$), such that larger values indicate broader spatial attention (shallower gradients). Each circular marker represents one participant. Marker size is scaled proportionally to the magnitude of the width measure ($1/\|\text{gradient}\|_2$), with larger markers indicating broader attentional spread / more flexible attention. Marker colour and the colour bar encode trait anxiety severity. The figure visualizes how cortical thickness predicts the width of attention, and how this relationship is modulated by trait anxiety severity. (F) Probability density plot of the null distribution of Pearson's correlation coefficients (*r*) from 2,000 random permutations, estimating the null hypothesis that the empirical (*r*) does not differ from zero for significance testing. The red dashed vertical line represents the empirical cross-validated (*r*). Thresholded statistical maps along with their respective colour bars are visualised using the inflated FSAverage template in the CAT12 toolbox in (A) and (C). Least squares trend lines (broken black, red [high anxiety subset], and blue [low anxiety subset]) are shown in (B) and (D), and prediction intervals (dashed black lines) are shown in E.

**Abbreviations**

**AAL3** – Automated Anatomical Labeling atlas (version 3)

**AMAP** – Adaptive Maximum a Posteriori

**ANCOVA** – Analysis of Covariance

**ANOVA** – Analysis of Variance

**CAT12** – Computational Anatomy Toolbox (version 12)

**CSF** – Cerebrospinal Fluid

**DAN** – Dorsal Attention Network

**EEG** – Electroencephalography

**FE** – Fear condition (Fearful emotion)

**FOV** – Field of View

**FSAverage** – FreeSurfer Average Template

**FWE** – Family-Wise Error

**FWHM** – Full Width at Half Maximum

**GAD** – Generalized Anxiety Disorder

**GM** – Gray Matter

**GMV** – Gray Matter Volume

**IFG** – Inferior Frontal Gyrus

**IQR** – Image Quality Rating

**IPS** – Intraparietal Sulcus

**ISI** – Inter-Stimulus Interval

**KDEF** – Karolinska Directed Emotional Faces

**LOC** – Lateral Occipital Complex

**MNI** – Montreal Neurological Institute

**MRI** – Magnetic Resonance Imaging

**nIES** – Negative Inverse Efficiency Score

**PBT** – Projection-Based Thickness

**PVE** – Partial Volume Estimation

**rGMV** – Regional Gray Matter Volume

**RT** – Reaction Time

**SBM** – Surface-Based Morphometry

**SPM** – Statistical Parametric Mapping

**STAI** – State–Trait Anxiety Inventory

**STAI-T** – State–Trait Anxiety Inventory – Trait subscale

**STAI-S** – State–Trait Anxiety Inventory – State subscale

**TIV** – Total Intracranial Volume

**TR** – Repetition Time

**TE** – Echo Time

**VBM** – Voxel-Based Morphometry

**VWM** – Visual Working Memory

**V1** – Primary Visual Cortex

**Figure 1**

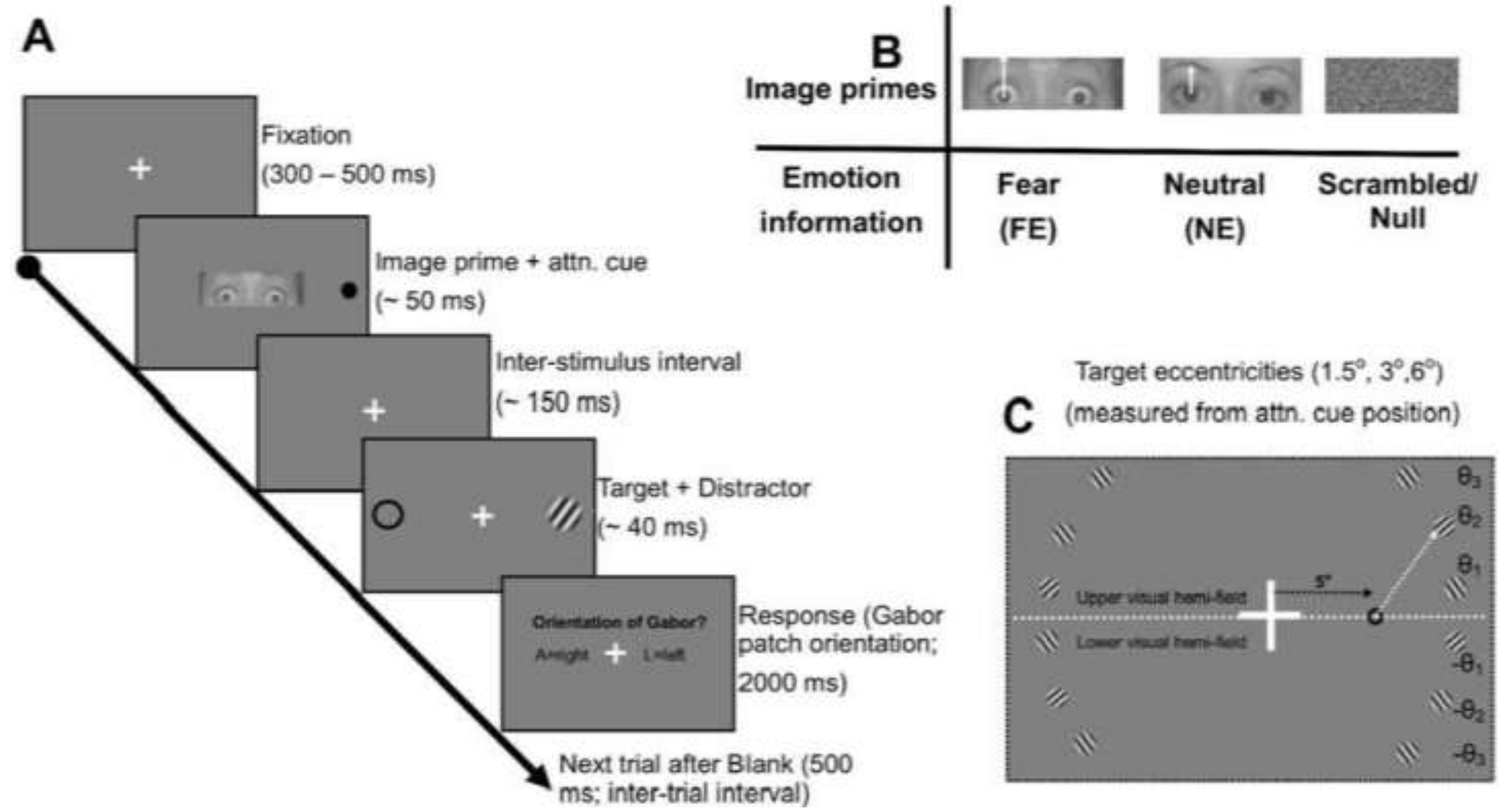
A
Fixation
(300 – 500 ms)
Image prime + attn. cue
(~ 50 ms)
Inter-stimulus interval
(~ 150 ms)
Target + Distractor
(~ 40 ms)
Orientation of Gabor?
Response (Gabor patch orientation; 2000 ms)
Next trial after Blank (500 ms; inter-trial interval)
B
Image primes
Emotion information
Fear (FE)
Neutral (NE)
Scrambled/ Null
C
Target eccentricities (1.5°, 3°,6°)
(measured from attn. cue position)
Upper visual hemi-field
Lower visual hemi-field
5°
θ3
θ2
θ1
-θ1
-θ2
-θ3

**Figure 2**

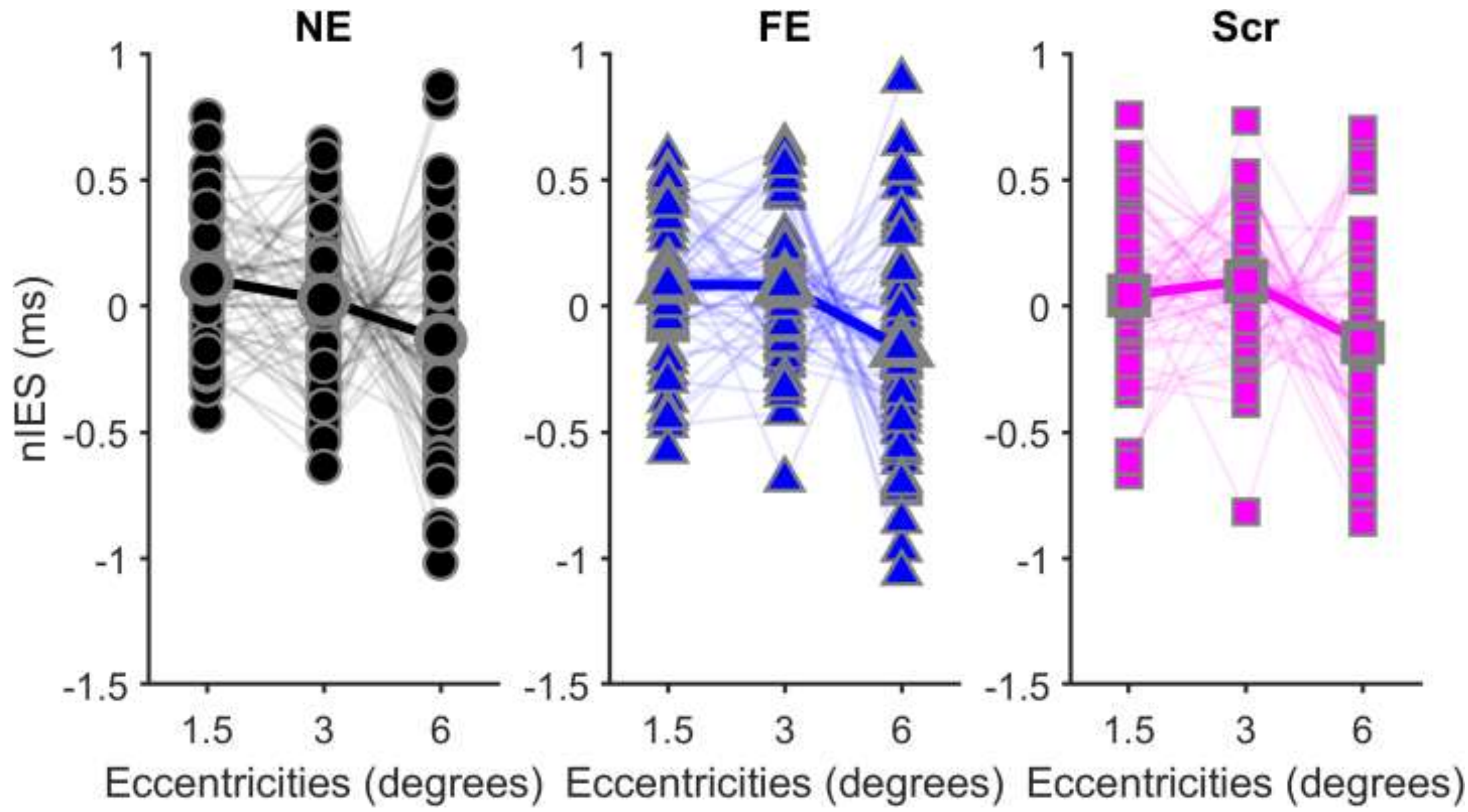

**Figure 3**

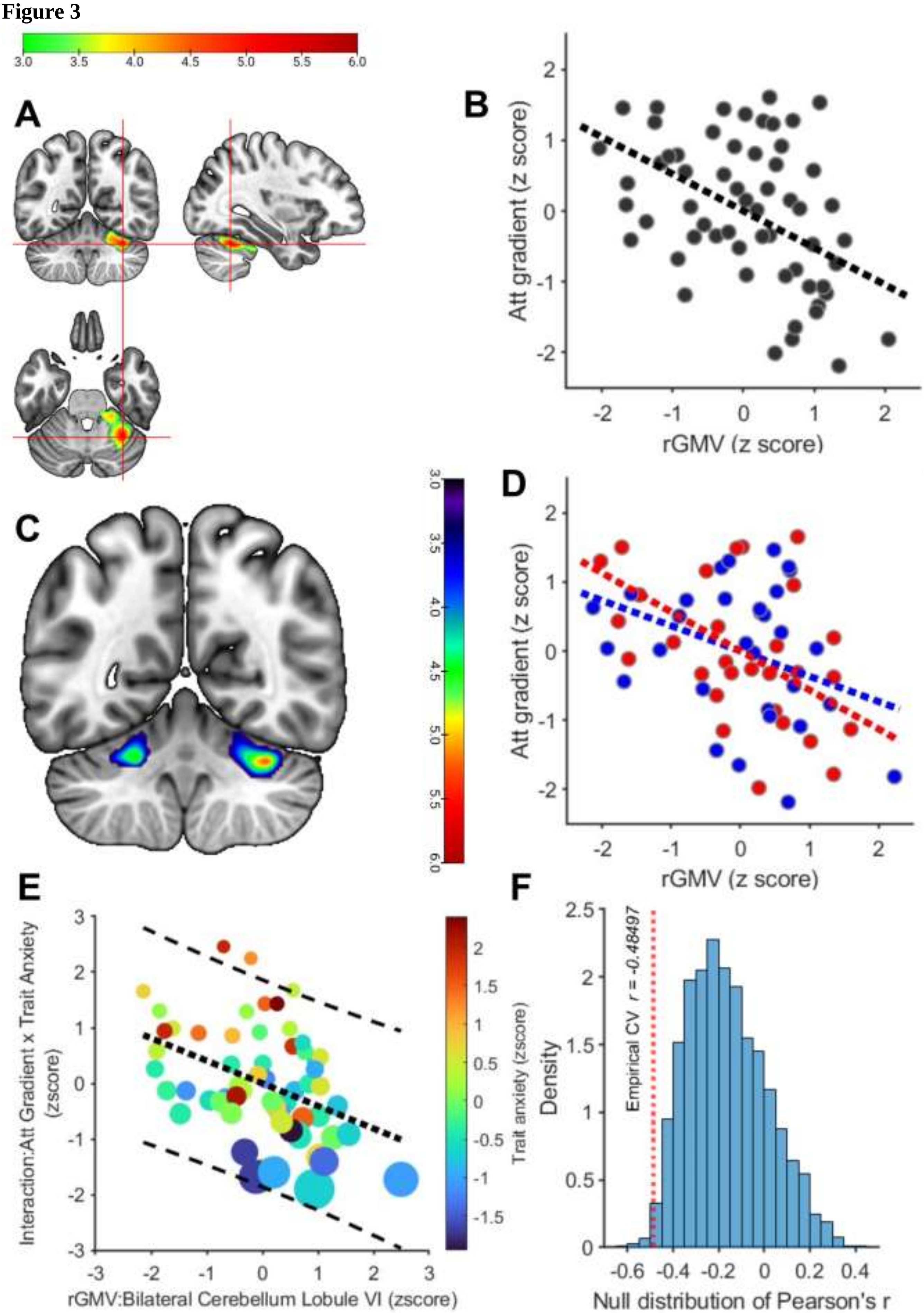
A
B
C
D
E
F
Att gradient (z score)
rGMV (z score)
Interaction:Att Gradient x Trait Anxiety (zscore)
rGMV:Bilateral Cerebellum Lobule VI (zscore)
Trait anxiety (zscore)
Density
Empirical CV r = -0.48497
Null distribution of Pearson's r

**Figure 4**

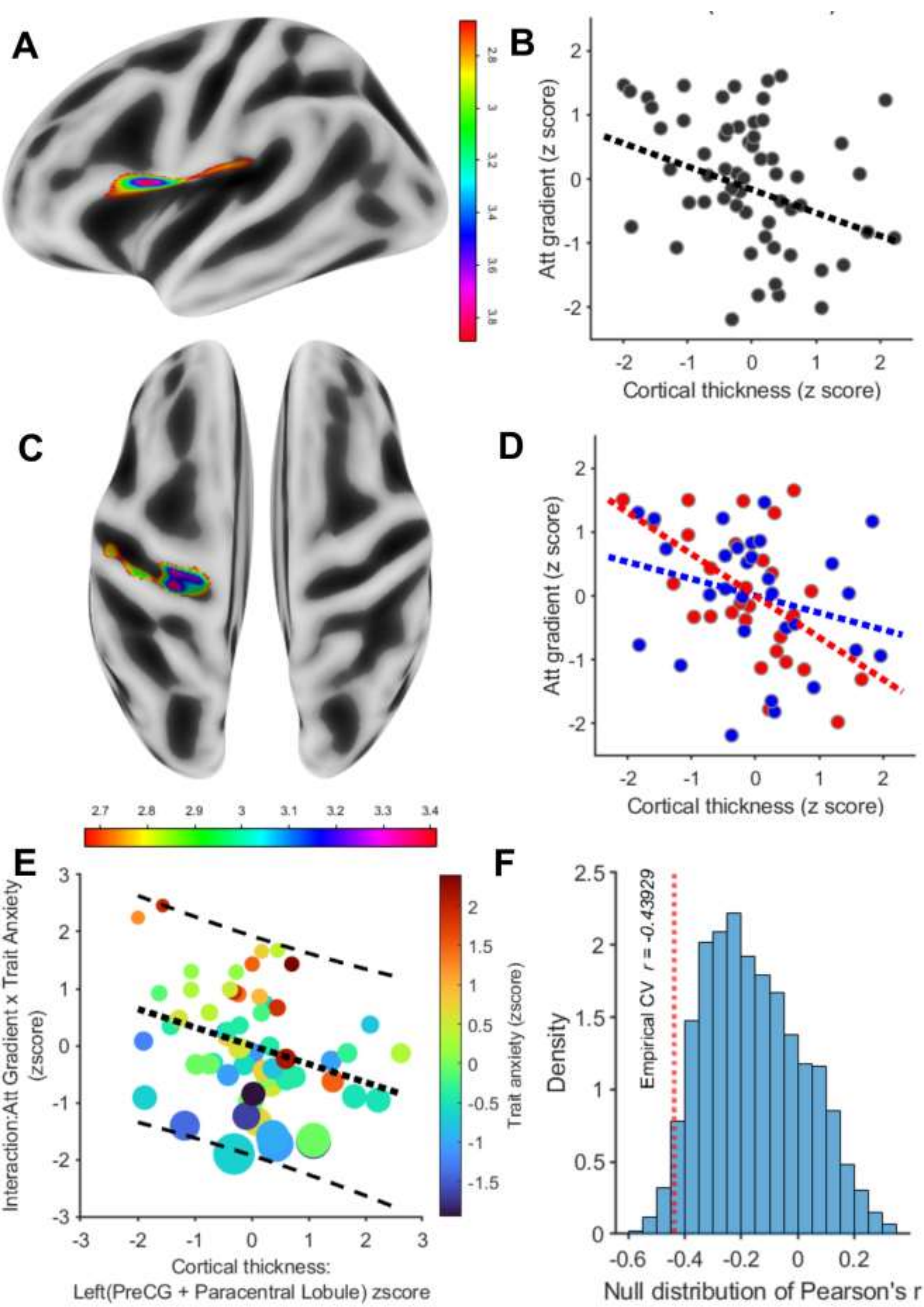
A
2.8
3
3.2
3.4
3.6
3.8
B
Att gradient (z score)
Cortical thickness (z score)
C
D
Att gradient (z score)
Cortical thickness (z score)
2.7
2.8
2.9
3
3.1
3.2
3.3
3.4
E
Interaction:Att Gradient x Trait Anxiety (zscore)
Trait anxiety (zscore)
Cortical thickness: Left(PreCG + Paracentral Lobule) zscore
F
Density
Empirical CV r = -0.43929
Null distribution of Pearson's r

# Supplementary Information

## 1. Supplementary Analysis 1

To ascertain the main effect of eccentricity on the attentional efficiency, the nIES values across three emotions were averaged at each eccentricity for each participant and entered into a one-way repeated-measures ANOVA with eccentricity condition as a within-subject factor (1.5º, 3º, 6º). Assumptions of normality were largely met (Lilliefors test; *p values of two groups* = 0.5) but not sphericity (Mauchly's test: $W = 0.73$, $p = 0.0001$). Thus, Greenhouse-Geisser corrected values are reported here. The main effect of Eccentricity was significant on the nIES scores ($F(2, 118) = 13.56$, $p = 0.000035$, partial $\eta^2 = 0.19$; Fig. 2), indicating that attentional efficiency at spatial loci fell off with increasing eccentricity from the centre.

## 2.Supplementary Analysis 2

Pearson's correlations indicated that neither the nIES values at each eccentricity (for the three emotional conditions) nor the single composite gradient (computed across the three emotion conditions) were significantly associated with trait anxiety (all uncorrected $p > .05$; see Supplementary Table 1)

**Supplementary Table 1**

| Predictor | Pearsons' *r* | *p* value | CI Lower | CI Upper |
|---|---|---|---|---|
| Neutral-1.5° | 0.05075771 | 0.70012981 | -0.2058197 | 0.30080544 |
| Neutral-3° | -0.0812386 | 0.53719837 | -0.3283891 | 0.17632342 |
| Neutral-6° | 0.02996738 | 0.82019486 | -0.2256746 | 0.28174814 |
| Fear-1.5° | -0.1026927 | 0.43492694 | -0.3475545 | 0.15528116 |
| Fear-3° | -0.1950193 | 0.13538112 | -0.4277612 | 0.06197439 |
| Fear-6° | 0.19830158 | 0.12879949 | -0.0585725 | 0.43054667 |
| Scrambled-1.5° | 0.14811209 | 0.25873769 | -0.1099477 | 0.38746451 |
| Scrambled-3° | 0.06748406 | 0.6084286 | -0.1896912 | 0.31599393 |
| Scrambled-6° | -0.1628074 | 0.2139105 | -0.400188 | 0.09504661 |
| CompositeGradient | 0.02323721 | 0.86011165 | -0.2320567 | 0.27553611 |

CI = 95% - Confidence Interval

To complement the analyses focusing on trait anxiety, we examined the relationships between state anxiety, trait anxiety, and the composite gradient measure. Trait and state anxiety were not significantly correlated in the present sample (Pearson's $r = -0.14$, $p = 0.292$, 95% CI [-0.38, 0.12]). Likewise, state anxiety was not significantly associated with the composite gradient measure (Pearson's $r = -0.25$, $p = 0.052$, 95% CI [-0.48, 0.00]).

**Whole-Brain Morphometric Correlates of Trait Anxiety:**

Whole-brain multiple regression analyses with trait anxiety (STAI-Trait) as the sole predictor of regional GMV after controlling for age and total intracranial volume as well as cortical thickness after controlling for age.

*Main Result: No clusters survived family-wise error (FWE) correction*.

*Results A [Positive correlation between Trait Anxiety and GMV – gray matter volume]*

All peak-level T values ≤ 4.38 and all cluster-level FWE-corrected *p*s ≥ 0.860

*Result B [Negative correlation between Trait Anxiety and GMV – gray matter volume]*

All peak-level T values ≤ 4.55 and all cluster-level FWE-corrected *p*s ≥ 0.650

*Result C [Positive Correlation between Trait Anxiety and SBM – cortical thickness]*

No voxels survived height threshold.

*Result D [Negative Correlation between Trait Anxiety and SBM – cortical thickness]*

All peak-level T values ≤ 2.80 and all cluster-level FWE-corrected *p*s ≥ 0.886

**Supplementary Fig. 1**

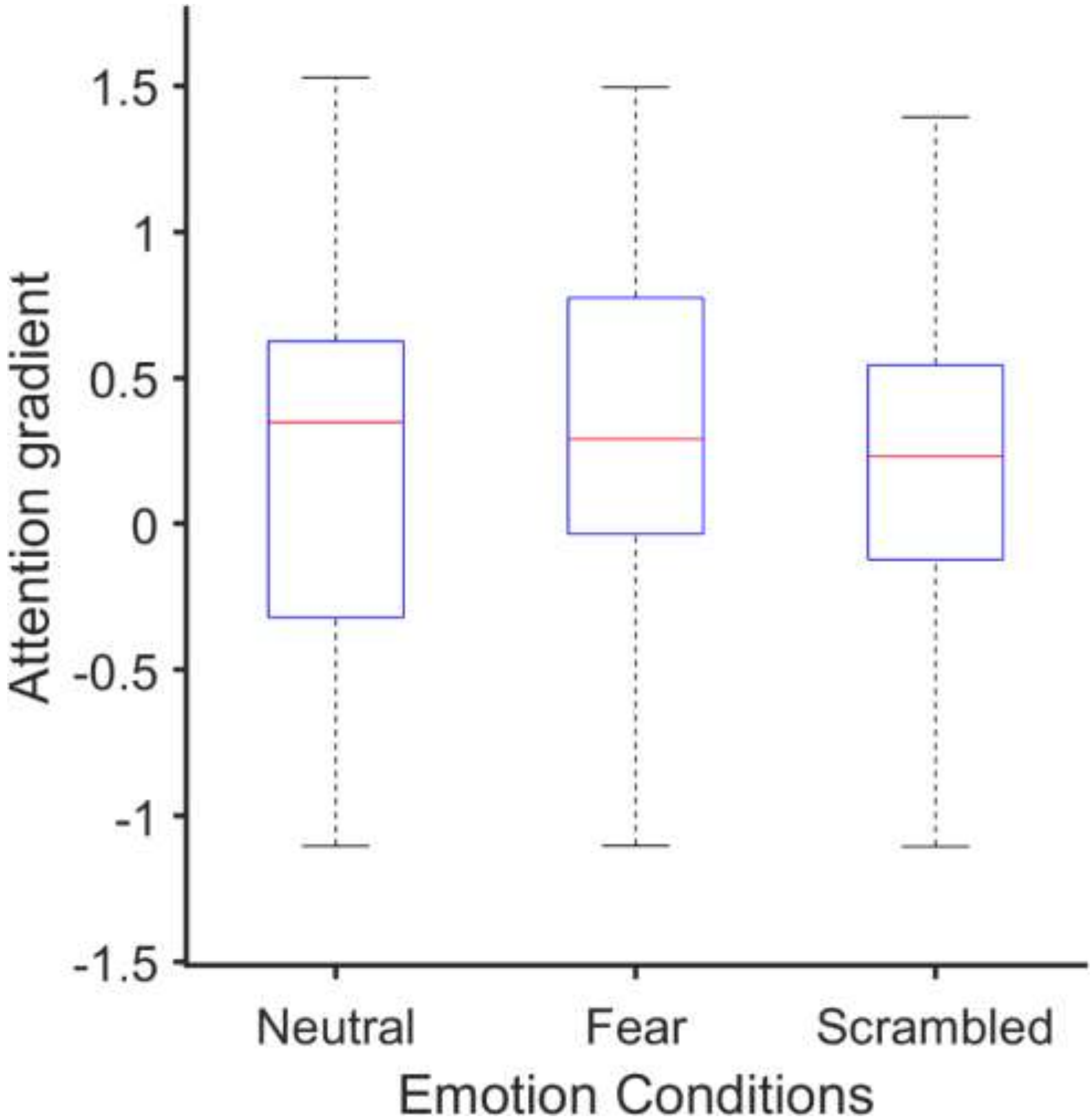

**Supplementary Fig. 1 Effects of partial Emotion pre-cues on Attention gradient**. Box and whisker plots showing the respective distributions and medians (red horizontal lines in the boxes) of the attention gradients pre-cued by three different partial emotions. Note that there was no statistically significant difference across the three emotions.

**Supplementary Fig. 2**

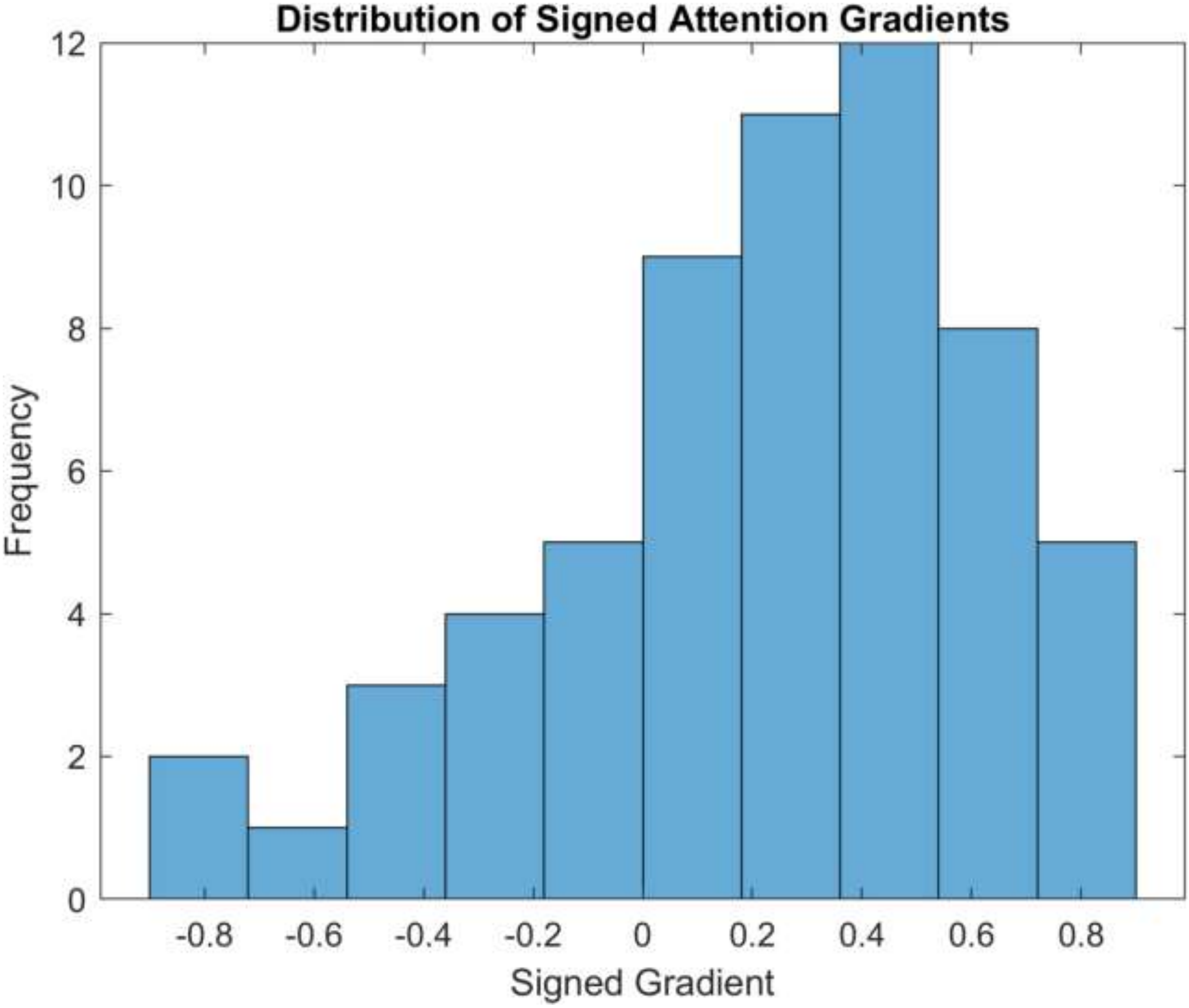

**Supplementary Fig. 2. Distribution of signed Near–Far attention gradients across participants**. The histogram shows the distribution of participant-level mean signed Near–Far gradients averaged across the three emotional conditions and used to derive the Euclidean norm composite metric. Positive values indicate relative prioritization of near (1.5°) compared to far (6°) locations, whereas negative values indicate relative prioritization of far compared to near locations. Although the Euclidean norm employed in the primary MRI analyses is direction-invariant, 75% of participants (45/60) exhibited positive signed gradients, indicating that near-space prioritization was the predominant attentional profile in the present sample.

**Supplementary Fig. 3**

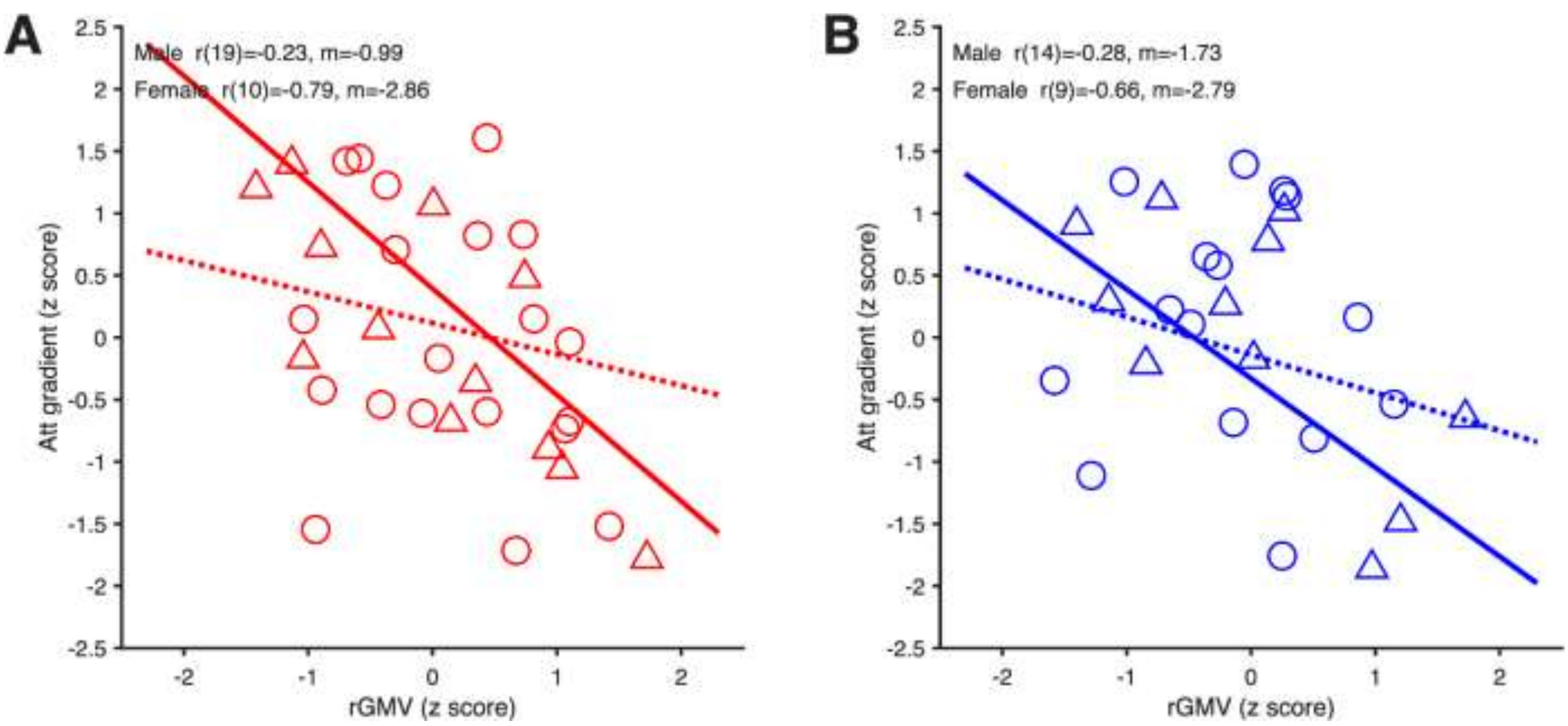

**Supplementary Fig. 3 Relationship of GMV with Attention Gradient**. (A) Negative correlation between attention gradient and GMV in bilateral cerebellar lobule VI in high anxiety subset males (red circles and broken trend line) and females (red triangles and solid trend line) (B) Negative correlation between attention gradient and GMV in bilateral cerebellar lobule VI in low anxiety subset males (blue circles and broken trend line) and females (blue triangles and solid trend line)

**Supplementary Fig. 4**

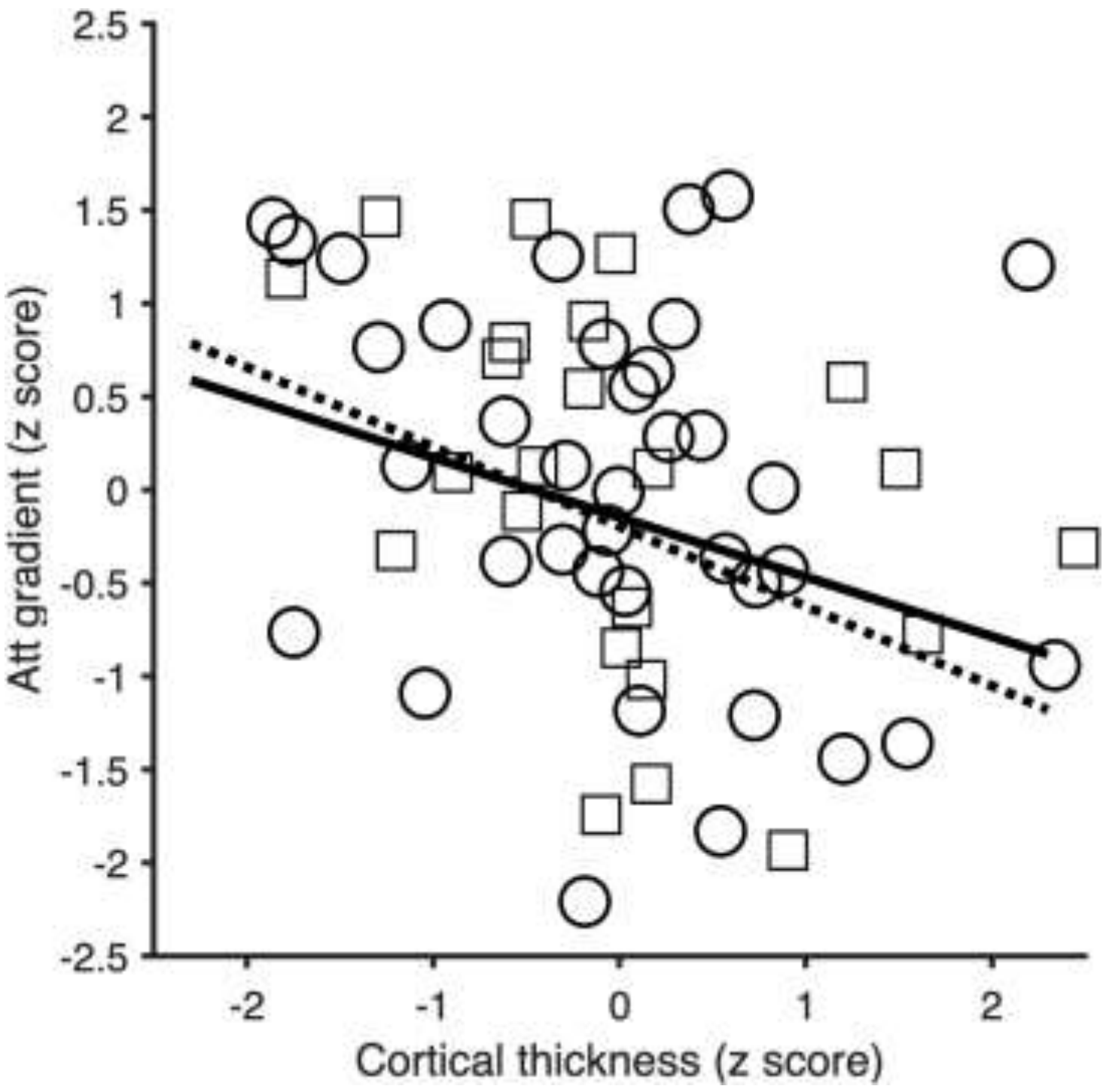

**Supplementary Fig. 4 Relationship of cortical thickness with Attention Gradient.** Negative correlation between attention gradient and cortical thickness in the left (precentral gyrus + paracentral lobule) in males (black circles and solid trend line) and females (black squares and broken trend line)

**3.Supplementary Analysis 3**

To support the primary analyses, we conducted a series of additional reliability, validity, and sensitivity analyses. These analyses evaluated the stability of the visuospatial attention gradient measure, its convergence with an alternative three-eccentricity slope-based metric, and the robustness of the reported neuroanatomical findings across two different methods of calculating the attentional gradients.

**3.1 Split-Half Reliability of the Composite Gradient Measure**

To evaluate the reliability of the composite Near–Far gradient as an individual-difference measure, the data were divided into two independent halves (first three sessions and last three sessions) and the Euclidean norm-based gradient metric was computed separately for each half. Because split-half correlations are based on shortened versions of the measure, the resulting correlation was adjusted using the Spearman–Brown formula to estimate the reliability of the full-length measure (BROWN, 1910; SPEARMAN, 1910). The split-half correlation was significant (Pearson's $r(58) = 0.56$, 95% CI [0.35, 0.71], $p < 0.001$), and the corresponding Spearman–Brown corrected reliability coefficient was 0.71. The Spearman–Brown corrected reliability coefficient exceeded 0.70, meeting the generally accepted empirical threshold for adequate internal consistency or stability of participant differences in the gradient measure (Hair *et al.*, 2019).

**3.2 Convergent Validity of the Near–Far Gradient Metric**

To determine whether the primary findings depended on the endpoint-based Near–Far gradient metric, we derived an alternative attentional gradient measure using all three eccentricities (1.5°, 3°, and 6°). For each participant and emotion condition, a simple linear regression was fitted to the nIES values across eccentricities. This was of the form: nIES = b0 + b1(Eccentricity) where b0 = intercept; b1 = slope coefficient. The resulting slope coefficient was extracted as an index of the rate of change in attentional efficiency (nIES) across space. The goodness-of-fit ($R^2$) of each participant-specific linear model was also computed to evaluate how well a linear gradient described the eccentricity profile. Analogous to the original analysis, the three emotion-specific slopes were combined using a Euclidean norm to generate a composite slope-based gradient metric. We then examined the correspondence between this alternative metric and the original Near–Far composite gradient using Pearson's correlations and 95% confidence intervals. Additional emotion-specific correlations were conducted to determine whether the endpoint-based Near–Far gradients and the corresponding three-point slope estimates captured similar individual differences in attentional deployment within each emotional condition (Neutral, Fear, Scrambled). Finally, summary statistics of the participant-specific $R^2$ were computed to characterize the overall quality of the linear fits.

We observed a strong positive association between the two measures when combined across three emotion conditions, which was significant (Pearson's $r(58) = 0.97$, 95% CI [0.95, 0.98], $p < 0.0001$; Supplementary Figure 5), indicating that both approaches captured highly similar individual differences in visuospatial attentional gradients. Consistent results were observed at the level of individual emotion conditions, where the endpoint-based gradients were significantly associated with their corresponding three-point slope estimates (Neutral: Pearson's $r(58) = -0.9927$, 95% CI [-0.9957, -0.9878], $p < 0.0001$; Fear: Pearson's $r(58) = -0.9938$, 95% CI [-0.9963, -0.9896], $p < 0.0001$; Scrambled: Pearson's $r(58) = -0.9911$, 95% CI [-0.9947, -0.9851], $p < 0.0001$). Note that the negative correlations observed at the emotion-specific level reflect the opposite sign conventions of the endpoint gradient (Near minus Far) and slope metrics, which capture the same underlying attentional profile in opposite directions. However, because the Euclidean norm removes directional information by squaring the emotion-specific values, these inverse relationships are transformed into a strong positive association between the corresponding composite metrics. To further evaluate the suitability of the slope-based approach, we examined the quality of the participant-specific linear fits across the three eccentricities. The linear model accounted for an appreciable proportion of variance in the eccentricity profiles (mean $R^2 = 0.66$, 95% CI [0.61, 0.70]).

Together, these findings provide evidence of convergent validity and suggest that the observed effects using the original Euclidean norm based composite gradient used in the reported findings are not dependent on the specific Near – Far endpoint-based quantification of the attentional gradient.

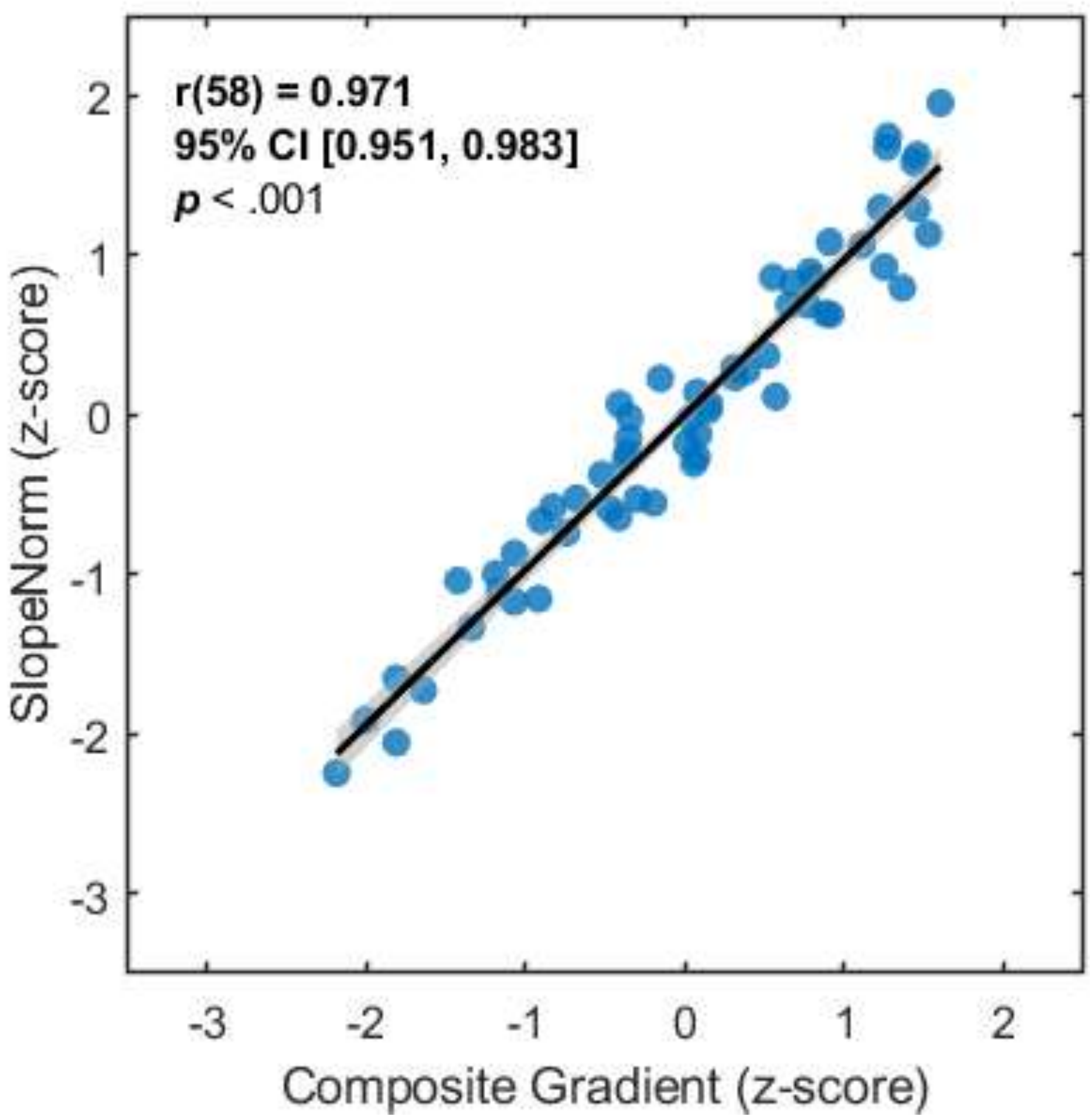

**Supplementary Figure 5**. **Convergent validity of the Near – Far attentional gradient metric**. Scatterplot showing the association between the original endpoint-based composite gradient (CompositeGradient) and the alternative three-eccentricity slope-based composite metric (SlopeNorm). The significant association between the two measures indicates that individual differences in attentional gradients are not dependent on the choice of the specific method used to quantify them.

**3.3 Sensitivity Analysis (Neuroimaging) Using a Three-Eccentricity Slope-Based Gradient Metric**

To determine whether the reported neuroanatomical findings depended on the endpoint-based Near–Far gradient metric, we repeated the principal VBM and SBM analyses using the alternative slope-based composite gradient derived from participant-specific linear slopes estimated across all three eccentricities. The resulting findings closely matched those obtained with the original gradient measure, indicating that the observed brain-behaviour relationships were robust to the method used to quantify the attentional gradient and were not driven by the specific choice of Near - Far endpoint-based method of gradient calculation. The same neuroanatomical regions identified using the original Near–Far gradient metric remained significant when the slope-based metric was used, although minor differences in cluster-forming thresholds were observed for the SBM analyses. The minor differences in SBM cluster-forming thresholds may be due to small differences in the variance captured by the alternative slope-based and endpoint-based gradient measures. Despite these quantitative differences, the same cortical regions were identified, indicating that the underlying neuroanatomical associations were robust to the method used to estimate the attentional gradient. Please see Supplementary Table 2 for details.

**Supplementary Table 2. Sensitivity Analysis Using a Three-Eccentricity Slope-Based Gradient Metric**

| **Analysis** | **Region** | **Peak MNI (x,y,z)** | **Peak T** | **Cluster Size** | **Cluster-Forming Threshold** | **FWE-corrected p** |
|---|---|---|---|---|---|---|
| VBM: Main Effect – Slope attention gradient | Cerebellar Lobule VI | 33, -56, -28 | 5.41 | 14 | $p < 0.001$ | 0.019 |
| SBM: Main Effect- Slope attention gradient | Front Inf Tri Left<br><br>Left Precentral Gyrus | -41, 22, 22<br><br>-46, 0, 35 | 3.92 | 528 | $p < 0.006$ | 0.034 |
| VBM : Trait Anxiety × Slope attention gradient | Right Cerebellum Lobule VI<br><br>Left Cerebellum Lobule VI | 33, -57, -28<br><br>-21, -60, -26 | 5.32<br><br>4.83 | 1964<br><br>871 | $p < 0.001$<br><br>$p < 0.001$ | < 0.001<br><br>0.022 |
| SBM: Trait Anxiety × Slope attention gradient | Left Precentral Gyrus<br><br>Left Paracentral Lobule<br><br>Left Precentral Gyrus | -23, -24, 66<br><br>-13, -21, 72<br><br>-38, -21, 61 | 3.32 | 764 | $p < 0.01$ | 0.016 |